\documentstyle[11pt]{article}
\begin{document}
\textwidth=6.0in
\textheight=20cm
\headheight=0cm
\setlength{\baselineskip}{0.8cm}
\fnsymbol{footnote}
\begin{center}
\ \vspace{3cm} . \\  
{\Large \bf   Geometry and Topology of \\ \ \\
     \huge  \bf         SOLITONS\\
               }
\vspace{3cm} \ 
{\bf \Large N. RIAZI }\\  \  \\
{\bf \large Department of Physics \\ Shiraz University \\
Shiraz 71454, Iran}\\ and      \\
{\bf \large Department of Physics and Astronomy \\University of Victoria \\
Victoria,  V8P 1A1\\ Canada  }
\end{center}
\vspace{1cm} \ 
\copyright N. Riazi, 1999.\\
\bf \large
\newpage
\pagestyle{headings}
\begin{center}
{\bf \Large Contents}\\ \ \\
\end{center}
\bf 
\begin{tabbing}
1. Introduction \ \ \ \ \ \ \ \ \ \ \ \ \ \ \ \ \ \ \ \ \ \ \ \ \ \ \ \ \ \ \ \ \ \ \ \ \ \ \ \ \ \ \ \ \ \ \ \ \ \ \ \ \ \ \= 3\\
2. Forms, fibers, and bundles  \> 3\\ 
3. Smooth maps and winding numbers  \> 9\\
4. Gauge fields as connections on principal bundles  \> 10\\
5. Yang-Mills field   \> 14 \\
6. Self-duality, instantons, and monopoles   \> 16 \\
7. Topological currents  \> 20 \\
8. Cohomology and electromagnetism  \> 24 \\
9. Homotopy groups and cosmic strings  \> 27 \\
10. Characteristic classes  \> 30 \\
11. Differential geometry and Riemannian manifolds \> 32 \\
12. Two-dimensional ferromagnet  \> 35 \\
13. Instantons in the $CP_N$ model  \> 37 \\
14. Skyrme model  \> 38 \\
15. Solitons and noncommutative geometry  \> 40
\\ \ \\
Acknowledgements  \> 43
\\ \ \\
References  \> 44
\end{tabbing}
\ \vspace{2cm} \ 
\begin{center}
{\large Abstract \\ \ \\}
\end{center}
\small
\bf
Basic concepts and definitions in differential geometry and topology 
which are important in the theory of solitons and instantons are 
reviewed. Many examples from soliton theory are discussed briefly, in 
order to highlight the application of various geometrical concepts and 
techniques. \newpage 
\ \\
{\bf \Large 1. Introduction }\\

Recent developments in soliton theory have been associated with frequent 
application of geometrical and topological ideas which provide an elegant interpretation
of many soliton properties. The mathematical methods of differential 
geometry and topology are very abstract and rigorous which make them 
hard to grasp by a physics-oriented scientist. In this writing, I have 
tried to provide an interdisciplinary and informal introduction to 
those topics in differential geometry and topology which have proven 
important in soliton theory. A background knowledge of
tensor calculus and field theory is assumed throughout this review.\\ \ \\
{\Large 2. Forms, fibers, and bundles }
\\

An $n-$dimensional  \emph{ manifold} is a space which behaves locally  like $R^n$.
In a similar manner, complex manifolds can be defined which are locally
similar to $C^n$. A circle  is a simple example of a one-dimensional 
manifold while a figure like $\LARGE +$ is not a manifold because of 
its behavior at the junction point. \emph{ Compact manifolds} have a finite 
volume.  An immediate example is the $n-$dimensional sphere $S^n$
 ( $S^o$ which contains only two points can be included in this 
 definition ).
In contrast, $R^n$ is an example of a noncompact manifold. 

\emph{ Group manifolds} are spaces constructed by the free parameters which 
specify the elements of a group. For example, 
the group manifold of $Z_2$ is $S^o$ while that of $U(1)$ is $S^1$, 
that of $SU(2)$ is $S^3$, that of $SO(3)$ is $SU(2)/Z_2$ or $P_3(R)$, etc.

Consider a real vector $E$ space on the $n-$dimensional manifold ${\mathcal M}$ 
( we are mainly concerned with $R^n$ and the Minkowski spacetime ).
\emph{ 1-forms} are linear mappings 
from $E$ to $R$:
\begin{equation}
\omega (\alpha u +\beta v ) =\alpha \omega (u) +\beta \omega (v)
\end{equation}
where $\omega $ is a 1-form , $\alpha , \beta \in R$, and 
$u,v \in E$. A familiar 
1-form in classical mechanics is the work 1-form $\vec{F} . d \vec{x}$. 
\emph{ $p-$forms}  $\omega ( u_1 , ..., u_p )$ are $p$-linear antisymmetric mappings from 
$p$-vectors $E\times ... \times E $ ( $p$-times )  to $R$: 
\begin{equation}
\omega ( u_1, ..., \alpha u'_j+\beta u''_j, ..., u_p ) = \alpha \omega 
( u_1, ... , u_j', ... , u_p ) 
+ \beta \omega ( u_1, ... , u_j'', ..., u_p ), 
\end{equation}
with
\begin{equation}
\omega ( u_{i_1}, ... , u_{i_p} ) = sgn (\pi ) \omega ( u_1, ... , u_p )
\end{equation}
where $sgn (\pi )$ =+1(-1) if the permutation 
$\pi : (1,...,p) \rightarrow ( i_1, ... , i_p ) $ 
is even ( odd ). Consider the $p-$form 
$\omega$ and the $q-$form $\eta$ such that $p+q$ is less than or equal to the dimension 
of the vector space $E$. The \emph{ exterior or wedge product} $\omega \wedge \eta $ is a $(p+q)-$form 
such that 
\begin{equation}
\eta \wedge \omega =(-1)^{pq} \omega \wedge \eta 
\end{equation}
This product is distributive with respect to addition, and 
associative: 
\[
\omega \wedge (\eta +\zeta )=\omega \wedge \eta +\omega \wedge \zeta,
\]
and
\begin{equation}
\omega \wedge (\eta \wedge \zeta ) = (\omega \wedge \eta ) \wedge \zeta .
\end{equation}

The space of all tangent vectors to a manifold ${\mathcal M}$ at the point 
$x$ is called the {\it tangent space } and is denoted by $T_x {\mathcal M}$. This space
has the same dimension $n$ as the manifold ${\mathcal M}$. The union of all such tangent
spaces form the \emph{tangent bundle} $T{\mathcal M}$ and is a manifold of dimension $2n$.

While $\{ \frac{\partial}{\partial x^i}\}$ forms a basis of the tangent space 
$T_x {\mathcal M}$,  the dual basis $\{ dx^i\}$ forms a basis of the 
so-called \emph{ cotangent space} $T^*_x ( {\mathcal M})$.  The inner 
product of these two bases satisfy 
\begin{equation}
( \frac{\partial}{\partial x^i}| dx^j ) =\delta^j_i
\end{equation}

\emph{ Differential $p-$forms} are $p-$forms on the tangent space $T{\mathcal M}$. 
Consider, for example, a real function $F(x,y,z)$ 
on the three dimensional Euclidean space $R^3$. $F$ is in fact a 
0-form, while $dF=\partial_i F dx^i$ is a differential 1-form, 
similar to the work form. The flux 2-form 
\begin{equation}
\Phi =\Phi_x dy\wedge dz +\Phi_y dz\wedge dx +\Phi_z dx\wedge dy 
\end{equation}
is another example of a differential form. In Minkowski spacetime, 
the electromagnetic potentials form a 1-form $A$ with components $A_\mu$ \
( $\mu = 0,1,2,3 $ ). 
The components of a $p-$form coincide with those of an anti-symmetric 
covariant tensor of rank p for $p>1$. For $p=0$, the 0-form transforms 
like a scalar and $p=1$ forms correspond to covariant vectors.

The tensor
product $\omega \otimes \eta$ of the $p-$form $\omega$ with the $q-$form $\eta$ 
does not make a $(p+q)-$form since $\omega \otimes \eta$ is not antisymmetric
with respect to all of its components. The wedge product $\omega \wedge \eta$
is defined in such a way to preserve the antisymmetry. For example, the 
wedge product of two 1-forms $\eta$ and $\omega$ satisfies $\omega \wedge
\eta =\omega \otimes \eta -\eta \otimes \omega$.  1-forms, therefore, 
anticommute ( $\omega\wedge \eta =-\eta \wedge \omega$ ). 

The exterior derivative of a $p-$form $\omega$ is 
defined in such a way to lead to a $(p+1)$-form. Consider, for example, 
the 1-form $A$ on $M$:
\begin{equation}
A=A_o dt+A_1 dx +A_2 dy +A_3 dz =A_\mu dx^\mu .
\end{equation}
We have
\[
dA =dA_o \wedge dt + dA_1 \wedge dx + dA_2 \wedge dy + dA_3 \wedge dz
=dA_\mu \wedge dx^\mu 
\]
\[
=(\partial_\alpha A_\mu dx^\alpha ) \wedge dx^\mu =\partial_\alpha A_\mu 
dx^\alpha \wedge dx^\mu 
\]
\begin{equation} 
= (\partial_y A_3 -\partial_z A_2 ) 
dy\wedge dz +... 
\end{equation}

The \emph{ Hodge operation} * on a $p-$form $\omega$ produces an $(n-p)$-form 
$\ ^*\omega$ according to 
\begin{equation}
(*\omega )_{ij...} =\frac{1}{p!}\epsilon_{lm...ij...} \omega^{lm...}
\end{equation}
where $\epsilon_{lm...}$ ( $n$ indices ) is the totally antisymmetric tensor 
of the $n-$dimensional space. Note that * is defined in terms of components
( the components of $p-$forms have p indices ). 
The antisymmetry of differential forms implies
\begin{equation}
dd \omega =0
\end{equation}
for any $p-$form $\omega$. It can also be shown that 
\begin{equation}
d(\omega \wedge \eta ) =d\omega \wedge \eta +(-1)^p \omega \wedge d\eta
\end{equation}
for any $p-$form $\omega$ and $q-$form $\eta$. The $p-$form $\omega$ is called 
\emph{closed} (globally) if $d\omega =0$ and \emph{exact}
if $\omega =d\eta$ where $\eta$ is a $(p-1)$-form.
Note that all exact forms are closed, but the 
converse is not always true. The reader recalls that not all vector 
fields can be written as the gradients of scalar functions.
  
The electromagnetic field 2-form $F$ defined by 
\begin{equation}
F=dA=\frac{1}{2} F_{\mu \nu} dx^\mu \wedge dx^\nu 
\end{equation}
is an example of an exact form. 
The action for the free Maxwell field can be written in the 
form 
\begin{equation}
-\int \frac{1}{2}\ ^*dA\wedge dA.
\end{equation}
Note that if $F_{\alpha \beta}$ are the components of a two-form $F$, 
$(dF)_{\alpha\beta\gamma}=\partial_\alpha F_{\beta\gamma} +
\partial_\beta F_{\gamma\alpha}+\partial_\gamma F_{\alpha\beta}$, which 
is antisymmetric with respect to all indices.
It can also be shown that if $\omega$ is closed and $\eta$ is exact, $\omega \wedge \eta$ will be 
exact. The reader can verify these two statements as exercises.

The well-known identities $\vec{\nabla}\times \vec{\nabla} 
\phi =0$ and $\vec{\nabla}. \vec{\nabla} \times
\vec{F} =0$ in vector analysis follow from the generalized 
identity $dd =0$. Consider, for example, the 1-form 
\begin{equation}
\omega = u_i dx^i
\end{equation}
in the three dimensional Euclidean space where $u_i $ ( $i=1,2,3$ ) are 
functions of $x^i$ ( $x^1=x,\ x^2=y,\ x^3=z  $ ). The 
exterior derivative of this 1-form yields $\vec{\nabla} \times 
\vec{u}$:
\[
d\omega = d( u_i dx^i ) = \frac{\partial u_i}{\partial x^j} 
dx^j \wedge dx^i 
\]  
\begin{equation}
= \frac{1}{2} ( \frac{\partial u_i}{
\partial x^j} -\frac{\partial u_j}{\partial x^i} ) dx^j\wedge dx^i
=\frac{1}{2} \epsilon_{kji} ( \vec{\nabla} \times
\vec{u} )_k dx^j\wedge dx^i
\end{equation}
and 
\begin{equation}
dd\omega =\frac{1}{2} ( \frac{\partial^2 u_i}{\partial x^k \partial x^j}
-\frac{\partial^2 u_j}{\partial x^k \partial x^i} ) dx^k\wedge dx^j\wedge
dx^i =0
\end{equation}
The last identity follows from the symmetry of $\frac{\partial^2}{
\partial x^k\partial x^j}$ and $\frac{\partial^2}{\partial x^k\partial
x^i} $ and antisymmetry of $dx^k\wedge dx^j$ and $dx^k \wedge dx^i$. 
 
A $p-$form which can be expressed as $\omega =d\phi_1\wedge...\wedge
d\phi_p$ where $\phi_1,...\phi_p$ are scalar fields, is called \emph{ 
simple}. A simple form  is both exact and closed.  Note that $\omega$ can
be written as $\omega =d(\phi_1 d\phi_2 \wedge ... \wedge d\phi_p)$.

The \emph{ co-differential } of the $p-$form $\omega$ is defined as 
\begin{equation}
\delta \omega =(-1)^{p(n-p+1)} \ ^*d^*\omega
\end{equation}
for Euclidean metrics. The definition for Minkowskian metrics differ by 
a - sign. Note that if $\omega$ is a $p-$form, $\delta \omega$ is 
a $(p-1)-$form. Also $\delta \delta =\pm \ ^* d^{2*} =0$. The 
Laplacian operator is obtained by forming $\triangle = - (\delta d +
d\delta )$, which in a Minkowski spacetime becomes the  d'Alembertian 
operator. A differential form $\omega$ is called \emph{ harmonic} if $\triangle 
\omega =0$. 
The inhomogeneous Maxwell equations can also be written in the form 
$\delta F =J$ and the  conservation of electric current as $\delta J=0$.

 Integration of $p-$forms over the space of interest ( or part of it ) is 
of great importance. In $R^n$, $dx^1\wedge ... \wedge dx^n=dx^1...dx^n$ provides the 
volume $n-$form. 
For a general metric, the volume form is given by $\sqrt{|g|}dx^1...dx^n$,
where $g$ is the determinant of the metric tensor.
\emph{Stokes's theorem} reads
\begin{equation}
\int_V d\omega =\int_{\partial V} \omega
\end{equation}
where the LHS integration is over the submanifold $V$ with 
the boundary $\partial V$. The manifold
over which the integration is performed ( and hence its boundary ) 
is assumed to be \emph{orientable}. The existence of a volume form on the 
manifold of interest guarantees its orientability. \emph{Klein bottle} and 
\emph{M\"{o}bius strip} are examples of nonorientable manifolds.
Stokes's
theorem leads to the more common Stokes and divergence theorems in vector
analysis.  

Consider a 3-dimensional spacelike  volume $V$ with the boundary $\partial V$. The magnetic 
flux through $\partial V$ is given by $\int_{\partial V} 
F$, while the electric flux is given by $\int_{\partial V} \ ^*F$
which vanishes in ordinary electromagnetism.
The electric charge contained in $V$ is given by $-\int_{V} \ ^* J$.
The Gauss theorem can therefore be written as
\[
\int_{\partial V}\ ^*F=-\int_V \ ^*J.
\]

The following results are relevant to our discussion: 
\\
\begin{itemize}
\item{Any $k-$dimensional regular domain  
$X$ of a manifold ${\mathcal M}$ ( see section 3 ) has 
 a boundary $\partial X$ which is a $(k-1)$-dimensional compact 
 manifold. $\partial X$ itself has no boundary, i.e. $\partial \partial X
 =0$. This is called \emph{ Cartan's lemma }. }
\item{If ${\mathcal M}$ is a simply connected manifold ( see 
section 9 ), then $\oint_C \omega =0$ 
for an arbitrary closed 1-form $\omega$. C is a closed curve in ${\mathcal M}$.}
\item{If ${\mathcal M}$ is simply connected, then all closed 1-forms $\omega$ on
${\mathcal M}$ are exact. }
\item{Consider a compact, oriented, $p-$dimensional submanifold $X$ of 
the simply connected manifold ${\mathcal M}$. Then, for any two cohomologous 
( see section 8 ) 
$p-$forms $\omega$ and $\eta$ on ${\mathcal M}$, $\int_X \omega =\int_X\eta$.}
\item{ A closed $p-$form $\omega$ on $S^p$ is exact if and only if $\int_{S^p} \omega
 =0$.} 
\end{itemize}

Note that any singularities in the fields described by 
 differential forms must be treated as holes in the 
 manifold on which the forms are defined. For example, consider $d\theta$
 on the $xy$-plane where $\theta =arctan (y/x)$:
 \begin{equation}
 d\theta =\frac{-y}{x^2+y^2} dx +\frac{x}{x^2+y^2} dy
 \end{equation}
 This form is closed but not exact. While every closed $p-$form ( $p>0$ ) 
 on $R^n$ is exact, (20) is not exact because of the singularity at $x=y=0$ 
 which introduces a hole in $R^2$. In other words, the manifold on 
 which (20) is defined is $R^2-\{ 0 \}$. Any closed 1-form $\omega$ on 
 $R^2-\{ 0\}$ integrates on a smooth closed curve $C$ according to 
 \begin{equation}
 \oint_C \omega = w(C,0) \int_{S^1} \omega
 \end{equation}
 where $w(C,0)$ is the  winding number  of $C$ with respect to 
 the origin ( see section 3 ).

The \emph {Hilbert product} of forms is defined as follows
\begin{equation}
<\omega |\eta > =\int_{\mathcal{M}} \ ^*\omega \wedge \eta 
=\int_{\mathcal{M}} \frac{1}{p!}\omega^{i_1...i_p} 
\eta_{i_1...i_p}\sqrt{|g|} dx^1...dx^n
\end{equation}
where $g$ is the determinant of the manifold's metric.
Note that $\omega$ and $\eta$ are both $k$-forms. For compact manifolds 
the Hilbert product is always well-defined. It can be easily shown 
that $<\ ^*\omega |\ ^*\eta > =\pm <\omega |\eta >$ 
depending on the signature being Euclidean (+) or Minkowskian (-). 

As an example, consider the action 
for a real Klein-Gordon field 
\[
S=\int (\frac{1}{2}\partial^\mu\phi\partial_\mu\phi -\frac{1}{2}
m^2\phi^2 )\sqrt{|g|} d^4x 
\] 
which can be written as
\begin{equation}
S=\frac{1}{2} <d\phi |d \phi> 
-\frac{1}{2}m^2 <\phi |\phi >=\int (\frac{1}{2} \ ^*d\phi \wedge 
d\phi -\frac{1}{2} m^2 \ ^*\phi \wedge \phi ).
\end{equation}

The tangent space $T_x {\mathcal M}$ is the prototype of a \emph{fiber}. 
In order to define a 
fiber more generally, consider the triplet $(T, B, \pi )$, where $T$ is 
the \emph{bundle space} ( or \emph{total space} ), $B$ is the base space and $\pi$ is a 
$C^\infty$ ( i.e. infinitely differentiable ) mapping from $T$ to $B$. 
Such a triplet is called a \emph{fibration}. An example is $({\mathcal M}\times V, {\mathcal M}, \pi)$
where ${\mathcal M}\times V$ is the product manifold of ${\mathcal M}$ with an arbitrary 
$m-$dimensional vector space $V$. This is called a {\it local vector bundle 
of rank $m$ }. The projection map $\pi$ is simply $\pi (x,u)=x$ in this 
case, where $x\in {\mathcal M}$ and $u\in V$. Part of the total space which
sits on top of the point $x\in B$ is $\pi ^{-1} (x)$ and is 
called a \emph{fiber}. The manifold $T$ can therefore be considered as a collection
of fibers or a {\it fiber bundle}. A $C^\infty$ mapping $f:B\rightarrow T$
such that $\pi \circ f =Id$ is called a \emph{section} of the fibration. 
Here, $Id_B$ is the identity map on $B$ ( $Id_B (x)=x$ ).\\

A fiber bundle is \emph{ locally trivializable} if it can be described by 
the product manifold $U_i\times F$ where $U_i$ is a neighborhood of the 
base manifold and $F$ is the fiber. Since local properties are not sufficient 
to describe the global topology of the bundle, a set of \emph{transition functions}
$\phi_{ij}$ are defined which specify how the fibers are related to each 
other in the overlapping region of the two neighborhoods $U_i$ and $U_j$. 
The transition function $\phi_{ij}$ is therefore defined as the mapping 
of the  fibers on $U_i$ to the fibers on  $U_j$ over the region 
$U_i \cap U_j$. For a trivial fiber bundle, all the transition functions 
can be the identity map. 
As we said before, $\{ \frac{\partial}{\partial x^\mu } \}$ is the standard basis for the local
frames of the tangent bundle $T {\mathcal M}$ while $\{ dx^\mu \}$ is the 
basis for the cotangent bundle $T^* {\mathcal M} $. Since 
$\frac{\partial}{\partial x^\mu} =\frac{\partial x'^\alpha }{\partial x^{\mu }}
\frac{\partial }{\partial x'^\alpha } $ and $dx'^{\mu } =\frac{\partial x'^{\mu }}
{\partial x^\alpha } dx^\alpha $, they define transition 
functions between two overlapping neighborhoods $U$ and $U'$ of the 
tangent and cotangent bundles, respectively.

In gauge field theories, we encounter a special type of fibration which 
is called a {\it principal fibration}. A principal fibration consists of 
a base manifold ${\mathcal M}$, the structure group $G$, and the manifold $P$ on 
which the Lie group $G$ acts. Like in an ordinary fibration, $\pi$ is 
a mapping from $P$ to ${\mathcal M}$. The principal fibration, therefore, is denoted 
by $(P, G, {\mathcal M}, \pi )$. If $G$ is a gauge group, a section in the 
principal fibration corresponds to the choice of a particular gauge. 
In the case that there exists a global $C^\infty$ section, the principal 
fibration is said to be {\it trivializable }. A trivializable fibration is 
isomorphic to $({\mathcal M}\times G, G, {\mathcal M}, pr_1)$. 

Any fiber bundle is trivial if the base space is contractible. Non-trivial
fiber bundles can therefore exist over topologically non-trivial base  
manifolds ( like $R^3 -{0}$, $S^1$, etc. ). 
Only for a trivial principal bundle  can one find a single gauge potential 
which is smooth over the entire base manifold. 

The fiber of a vector bundle is a linear vector space. The transition 
functions of a vector bundle are elements of the \emph{general linear group} 
of the vector space. Similarly, the transition functions of  a principal 
bundle are elements of $G$ acting  by left multiplication. 
The associated vector bundle $P X_\rho V$ is defined using the representation
$R(G)$ acting on the finite-dimensional vector space $V$. The \emph{associated 
vector bundle} is based on the equivalence relation 
\begin{equation}
(p, \rho (g)\circ v ) \simeq (p\circ g, v) \ \ \ \forall \ \ \ p\in P, \ \ \
v\in V, \ \ \ g\in G, \ \ \ and \ \ \ \rho \in R(G).
\end{equation}

The tangent space to the bundle space $T_p P$ can be decomposed into 
{\it vertical } and {\it horizontal} parts:
\begin{equation}
T_p P=H_p \oplus V_p
\end{equation}
The vertical part which corresponds to the action of $G$ maps into a single 
point of the base space. The horizontal part leads to the {\it connection} 
1-form which will be defined in the next section.

The space of spinors is also a vector space and \emph{spinor bundles} can be 
constructed in a similar way as the vector bundles. The corresponding 
principal bundle has the spin group with the \emph{Clifford algebra} as the 
tangent space at the identity element.
\\ \ \\
{\bf \Large 3. Smooth maps and winding numbers}
\\

Consider a smooth map from the $n-$dimensional differentiable manifold 
${\mathcal M}$ to the $m-$dimensional differentiable manifold ${\mathcal M}'$
\begin{equation}
f:{\mathcal M} \rightarrow {\mathcal M}'
\end{equation}
This mapping is called
\begin{itemize}
\item{\emph{surjective}, if $f({\mathcal M} )={\mathcal M}'$. In other
words, for all $y\in {\mathcal M}'$ there is at least one
element $x\in {\mathcal M}$ such that $f(x)=y$.}
\item{\emph{injective}, if for all $x_1,x_2\in {\mathcal M}$ where
$x_1\neq x_2$, we have $f(x_1)\neq f(x_2)$. In other words, distinct
points in ${\mathcal M}$ have distinct images.}
\item{\emph{bijective}, if it is both surjective and injective. 
The existence of a bijective mapping from ${\mathcal M}$ to 
${\mathcal M}'$ ensures that the points in these two spaces 
are in one-to-one correspondence.}
\end{itemize}
For details, the reader is referred to Long (1971).

Let $x^i$ ( $i=1,...,n$ ) and $y^j$ ( $j=1,...,m$ ) be coordinate systems 
in ${\mathcal M}$ and ${\mathcal M}'$, respectively. These coordinates
are related by the map $f$. We can expand $y^j (x^i)$ around a point 
$p$ in ${\mathcal M}$ with coordinates $x_o^i$ 
\begin{equation}
y^j(x_o^i +\Delta x^i ) =y^j(x_o^i ) +
\left( \frac{\partial y^j}{\partial x^i} \right)_p \Delta x^i +...
\end{equation}
Note that $[\frac{\partial y^j}{\partial x^i} ]$ is an $m\times n$
matrix called the \emph{ Jacobi matrix}. This matrix defines the appropriate
linear mapping from $T_p ({\mathcal M} )$ to $T_{f(p)} ({\mathcal M} ')$
(see Felsager, 1983). 

A well-defined transformation between the two tangent spaces requires the 
mapping to be a \emph{ diffeomorphism}, which in simple terms means that 
the map is bijective with an inverse $f^{-1}$ which is smooth. For $m=n$, 
the Jacobi matrix becomes a non-singular square matrix. If $m<n$ and 
$f$ is everywhere regular, the map is called an \emph{ immersion}, 
and $f( {\mathcal M} )$ is a submanifold of ${\mathcal M}'$. An \emph{embedding} 
is an immersion which is further required to be a \emph{ homeomorphism}. 
If $m>n$, and $f$ is everywhere regular, it is called a \emph{submersion}. 
We now assume that ${\mathcal M}$ and ${\mathcal M}'$ are both compact and 
orientable, and have the same dimension $n$. The \emph{ Brouwer degree} of 
the map $f$ is defined as 
\begin{equation}
deg(f)=\sum_{p_i}sgn |\frac{\partial y^j}{\partial x^i}|_{p_i}
\end{equation}
where $f(p_i)$ are the regular values in ${\mathcal M}'$. 
A point $x\in {\mathcal M}$ is said to be \emph{regular} if $f'(x ) 
\neq 0$, otherwise, it is a \emph{ critical point}. A point $\xi\in 
{\mathcal M}'$ is either the image of a regular point, the image of a 
critical point, or it is not the image of any point.  
To see what a regular value means, consider a map $f:S^1\rightarrow S^1$. 
In this example, 
the map $\xi =\theta$ is of degree $+1$, $\xi=-\theta $ is of degree 
$-1$, and the map $\xi =\frac{1}{\pi} (\theta -\pi )^2$ has a vanishing 
degree. Note also that  $\xi=n\theta$ is of degree $n$. 

The Brouwer degree of a map measures the net number of times $\mathcal {M}'$ is 
covered when all the points in ${\mathcal M}$ are swept once. The 
integer $deg(f)$ is also called \emph{ winding number}. The Brouwer degree 
vanishes for maps which are not surjective. 

\emph{ Brouwer's theorem} states that 
\begin{equation}
\int_{\mathcal M} f^*\omega =deg(f)\int_{{\mathcal M}'}\omega
\end{equation}
where $\omega$ is a $p-$form on ${\mathcal M}'$ and $f^*\omega$ is 
its pullback, 
which is a $p-$form on ${\mathcal M}$ and will be defined in the
next section. 
\\ \ \\
{\bf \Large 4. Gauge Fields as connections on principal bundles}\\

Consider a gauge group represented by $m\times m$ complex matrices 
acting on a complex $m-$dimensional vector space $V$. A vector
bundle can be constructed with fibers isomorphic to $V$. A connection on 
this vector bundle is a 1-form on the base space with values in $C(m\times m)$
 (the space of complex $m\times m$ matrices).
The connection 1-forms of $SU(N)$, for example, are anti-hermitian, traceless
$N\times N$ matrices and at the same time 1-forms on the base manifold 
( e.g. the Minkowski space ). For a smooth section $S$ of the principal bundle,
the covariant derivative is defined according to
\begin{equation}
DS=ds+A\wedge S
\end{equation}
where $A=A_\mu dx^\mu$ is the connection 1-form. Note that each 
$A_\mu$ is a complex $N\times N$ matrix, expandable in terms of the 
bases of $SU(N)$:
\begin{equation}
A_\mu = A^a_\mu T_a.
\end{equation}
In the case of $SU(2)$, $T_a=\tau_a$ ( a=1,2,3 ) are the Pauli matrices
\begin{equation}
\tau_1=\left( \begin{array}{ll} i & 0 \\ 0 & \mbox{$-i$} \end{array} \right), \ \ \ \ 
\tau_2=\left( \begin{array}{ll} 0 & 1 \\ -1 & 0 \end{array} \right), \ \ \ \ 
\tau_3=\left( \begin{array}{ll} 0 & i \\ i & 0 \end{array} \right).
\end{equation}

The Lie algebra of a group $G$ is the tangent space to the group manifold 
at the identity element $T_e  G$. The basis $\{T_a \}$ of this space 
obeys the algebra 
\begin{equation}
[T_a,T_b]=f_{abc}T_c
\end{equation}
where $f_{abc}$ are the \emph{structure constants} of the group. 

Unlike the exterior derivative of ordinary 1-forms for which we have 
$ddS=0$, 
$DDS$ does not always vanish. This quantity which is a 2-form leads in  a
natural way to the concept of curvature. Taking the covariant derivative of 
(31),
\begin{equation}
DDS=D(A\wedge S)=dA\wedge S + A\wedge DS = F\wedge S,
\end{equation}
where the curvature 2-form is 
\begin{equation} \label{bdl}
  F=dA + A\wedge A.
\end{equation}
It can be shown that the covariant derivative of $F$ vanishes
\begin{equation}
DF=dF+A\wedge F- F\wedge A =0,
\end{equation}
which is the generalized form of  the {\it Bianchi identity}. Like $A$, $F$ is also in 
the form of an $m\times m$ complex matrix. $F$ is called curvature because
it is related to the Gaussian curvature when a curved manifold is 
concerned. The Bianchi identity constitutes one of the basic equations 
governing the gauge field $A$. An element $g$ of the gauge group 
linearly transforms a vector $v\in V$
\begin{equation}
v\rightarrow gv.
\end{equation}
Such a transformation is associated with the following gauge transformations
on the connection and curvature:
\begin{equation}
A\rightarrow Ad(g^{-1})A +g^{-1}dg,
\end{equation}
and
\begin{equation}
F\rightarrow gFg^{-1},
\end{equation}
where $Ad(g^{-1})$ is the adjoint representation of $g^{-1}$. For 
a single point as the base space, the bundle space reduces to the Lie group $G$ and 
the covariant derivative of the connection 1-form vanishes
\begin{equation}
dA+A\wedge A =0.
\end{equation}
This equation is called the {\it Maurer-Cartan structure equation}. 
The \emph{Maurer-Cartan form} $g^{-1}dg$ belongs to the Lie algebra of the 
principal bundle. This form is invariant under the left action  of a
constant group element $g_o$:
\begin{equation}
(g')^{-1}dg'=(g_og)^{-1} d(g_o g) =g^{-1}dg
\end{equation}
This form can be parametrized as $g^{-1}dg=\phi_aT_a$ where 
$T_a$ satisfy the  algebra (34).
Using the identity $d(g^{-1}dg)=-g^{-1}dg\wedge g^{-1}dg$ we obtain 
\begin{equation}
d\phi_a +\frac{1}{2} f_{abc} \phi_a \wedge \phi_b =0
\end{equation}
As we move along a curve $x^\mu (\lambda )$ in the base manifold, a 
corresponding section  $S(\lambda )$ ( called a \emph{ lift} ) is traced in
the principal bundle according to
\begin{equation}
\frac{d}{d\lambda } =\dot{x}^\mu \frac{\partial }{\partial x^\mu} 
+\dot{S}\frac{\partial}{\partial S}
\end{equation}
The section $S$ is said to be parallel transported if 
\begin{equation}
\dot{S}_{ij} +A_{\mu , ik} \dot{x}^\mu S_{kj} =0.
\end{equation}
From (44) and (45) we obtain
\begin{equation}
\frac{d}{d\lambda} =\dot{x}^\mu ( \frac{\partial}{\partial x^\mu} -
A^a_\mu T^a_{ij} S_{jk} \frac{\partial}{\partial S_{ik}} ).
\end{equation}
The expression inside the parentheses gives the
\emph{ covariant derivative}  denoted by $D_\mu$:
\begin{equation}
D_\mu = \partial_\mu -A^a_\mu T_a
\end{equation}
where $T_a =T^a_{ij} S_{jk} \frac{\partial }{\partial S_{ik}}$. 
The basic motivation for defining the covariant derivative
is to modify $\partial_\mu$ in such a way that the
resulting quantity transforms covariantly under the
action of the group element $g$.
The components of the curvature 2-form ( $F^a_{\mu\nu}$ ) 
are related to $D_\mu$ according to
\begin{equation}
[D_\mu , D_\nu ] =-F^a_{\mu \nu} T_a ,
\end{equation}
or 
\begin{equation}
F^a_{\mu \nu} =\partial_\mu A^a_\nu -\partial_\nu A^a_\mu 
+f_{abc} A^b_\mu A^c_\nu .
\end{equation}
In the language of differential forms, 
\begin{equation}
\Omega =d\omega +\omega \wedge \omega =g^{-1} Fg
\end{equation}
where $\omega =g^{-1} Ag +g^{-1} dg$ is the connection 1-form. 
Equations (36) and (49) can be combined to obtain
\begin{equation}
F=dA +A\wedge A =\frac{1}{2} F^a_{\mu\nu}T^a dx^\mu\wedge dx^\nu .
\end{equation}

The inhomogeneous  equations governing the gauge field read
\begin{equation}
D^*F =\ ^*J
\end{equation}
where $J$ is the current 1-form. Conservation of the current $J$ 
results from the underlying gauge symmetry  ( Noether's theorem )
\begin{equation}
D^*J=0
\end{equation}
Note that it is the dual of the current 1-form which is covariantly closed. 
The source-free field equations can  be obtained from the following 
action:
\begin{equation}
{\mathcal A}=\int \frac{1}{4} F\wedge ^* F
\end{equation}
Note that $F\wedge \ ^*F $ is a 4-form proportional to the 
volume 4-form in the 4-dimensional Minkowski space. 

In the case of the 
abelian $U(1)$ gauge symmetry, the group manifold is  a circle 
parametrized by the angle $\theta$ ( $g=e^{i\theta}$ ). The group element 
$g$ acts on the vector space of complex functions $\phi$. The 
principal bundle is locally isomorphic to the 
product of an open set in the Minkowski space and $S^1$. The potential 
1-form $A=A_\mu dx^\mu$ has real components $A_\mu$ in this case. We also
have $F=dA=\frac{1}{2} F_{\alpha \beta} dx^\alpha \wedge dx^\beta$ with 
components as a covariant antisymmetric tensor
\begin{equation}
F_{\alpha \beta} =\partial_\alpha A_\beta -\partial_\beta A_\alpha
\end{equation}
Note that $A\wedge A=0$, since in this case the potential 1-form is 
not matrix valued and $A_\mu A_\nu dx^\mu\wedge dx^\nu$ vanishes due to the 
symmetry of $A_\mu A_\nu$ and antisymmetry of $dx^\mu \wedge dx^\nu$. 
Also note that $A\wedge F=0$ and the 
Bianchi identity (37) translates into the following tensorial relation
\begin{equation}
\partial_\alpha F_{\beta \gamma} +\partial_\beta F_{\gamma \alpha} 
+\partial_\gamma F_{\alpha \beta } =0
\end{equation}
while (52) becomes
\begin{equation}
\partial_\beta F^{\alpha \beta} =J^\alpha
\end{equation}
Equations (56) and (57) constitute the complete Maxwell equations.

The $U(1)$ gauge transformation $\phi\rightarrow e^{i\theta}\phi$ leads 
to 
\begin{equation}
A\rightarrow A+d\theta, \ \ \ \ or \ \ \ \ A_\mu \rightarrow A_\mu +\partial
_\mu \theta 
\end{equation}
while $F$ remains gauge invariant. A section of the principal bundle 
corresponds to the selection of a  particular gauge $\theta (x^\mu)$. 

Maxwell's equations imply the
nonexistance of magnetic monopoles. The possibility of having
magnetic monopoles and its implications has been extensively 
studied in the literature ( see Goddard and Olive, 1978 and also
Horva'thy, 1988 for an introductory exposition of the subject). 

In the presence of magnetic charges, the homogeneous Maxwell equation 
is modified as 
\begin{equation}
dF=-^*K
\end{equation}
where $K$ is the magnetic current 1-form. Expressed in tensor
components, this equation reads 
\begin{equation}
\frac{1}{\sqrt{|g|}} \partial_\mu ( \sqrt{|g|}\ ^*F^{\mu\nu} ) =K^\nu
\end{equation}
or
\begin{equation}
\partial_\alpha F_{\beta \gamma} +\partial_\gamma F_{\alpha \beta}
+\partial_\beta F_{\gamma \alpha}=-\sqrt{|g|} \epsilon_{\alpha\beta
\gamma\delta} K^\delta
\end{equation}

A point-like monopole located at the
origin of spherical coordinates generates the following field
\begin{equation}
F=\frac{g}{4\pi} sin \theta d\theta \wedge d\phi
\end{equation}
where $g$ is the strength ( magnetic charge ) of the monopole. From 
$F=dA$ the 
potential 1-form can be chosen as
\begin{equation}
A=-\frac{g}{4\pi} cos \theta d\phi 
\end{equation}
The total magnetic flux across a sphere centered at the origin is 
\begin{equation}
\Phi_B =\oint_{S^2} F=\frac{g}{4\pi}\int^{2\pi}_{\phi=0}
\int^{\pi}_{\theta =0} sin\theta d\theta d\phi =g
\end{equation}
Note that the gauge potential (63) is \emph{ not} smooth everywhere. 
By computing the components of $A$ in Cartesian coordinates, it is 
easily seen that the potential diverges along the z-axis. Expressed in 
mathematical terms, this means that the corresponding bundle is not 
trivializable and a global section does not exist. 

The total angular momentum of the electromagnetic field produced by a 
static pair of a point charge $q$ and a magnetic monopole $g$ is given 
by
\begin{equation}
\vec{J} =\int \vec{x}\times (\epsilon_o \vec{E}\times\vec{B} ) d^3x
=\frac{qg}{4\pi} \hat{k}
\end{equation}
where $\hat{k}$ is a unit vector from $q$ to $g$. Using the quantum 
mechanical quantization $J=n\frac{\hbar}{2}$, we are led to an 
explanation for the quantization of the electric charge $q$ in the 
presence of a magnetic monopole. 
The Dirac monopole will be considered again in section 8.

Transition functions on principal bundles play the role of gauge 
transformations. Two fiber coordinates $\phi$ and $\phi '$ in $U\cap U'$ 
transform to each other by $\phi '=g\phi $ where $g$ is the transition 
function. Under this transformation, 
\begin{equation}
A'=g Ag^{-1} +g dg^{-1}, \ \ \ \omega '=\omega, \ \ \ 
F'=g Fg^{-1}, \ \ \ \ and \ \ \  \ \Omega ' =\Omega .
\end{equation}
In mathematical terms, $A$ and $F$ are called \emph{pullbacks} of $\omega$ and 
$\Omega$, respectively. 

The correspondence between the principal bundles and gauge fields can 
be summarized as follows

\begin{center}
Structure   group   $\leftrightarrow$   Gauge   group 
\\
Connection   pullback (A)  $\leftrightarrow$   Gauge   potential 
\\
Curvature   pullback  (F)   $\leftrightarrow$  Field   strength 
\\
Associated   vector   bundles $\psi$   $\leftrightarrow$ 
  Matter   fields  
\\
Transition   functions   $\leftrightarrow$  \ Gauge   transformations
\\
Sections   $\leftrightarrow$   Gauge   fixing   conditions 
\\
Maurer-Cartan   1-forms   $\leftrightarrow$   Pure   gauges 
\end{center}

The global topology of gauge fields become relevant in the path integral 
quantization approach. Path integral formalism works properly only for 
(++++)-signature spaces. This is why the Euclidean signature is 
of particular importance in the soliton and instanton methods.

For a detailed discussion of gauge theories and differential geometry, 
the reader is referred to Eguchi et al. ( 1980 ). 
\\ \ \\
{\bf \Large 5. Yang-Mills field}
\\

$SU(2)$ gauge theory was introduced in 1954 by Yang and Mills. The gauge
group of the Yang-Mills field is $SU(2)$.  Elements of this group are 
unitary $2\times 2$ matrices which 
operate on the two-component isospinors $\psi$. The covariant 
derivative $D_\mu $ is defined in such a way that $D_\mu \psi$ 
transform in the same way as $\psi$. According to equation  (47)
\begin{equation}
D_\mu =\partial_\mu -A^a_\mu \tau_a
\end{equation}
where $\tau_a$ are the Pauli matrices given in (33).
Note that this is a matrix-valued equation and $\partial_\mu$ is 
implicitly assumed to be  $I.\partial_\mu$ where $I$ is the 
$2\times 2$ unit matrix. The tangent space at the group identity $e=I$
defines the Lie algebra of the group. The algebra
of group generators $\tau_a$ is 
\begin{equation}
[ \tau_a , \tau_b ]=2i\epsilon_{abc} \tau_c
\end{equation}
where $\epsilon_{abc}$ is the totally antisymmetric tensor
in three dimensions. Note that 
according to (48) we have
\begin{equation}
[D_\mu , D_\nu ] \psi =iF_{\mu \nu} \psi 
\end{equation}
where
\begin{equation}
F_{\mu \nu } =\partial_\mu A_\nu -\partial_\nu A_\mu +i[ A_\mu , 
A_\nu ]
\end{equation}
which is the tensor version of equation ....(\ref{bdl})  % (36). Since 
$F_{\mu \nu} =iF^a_{\mu \nu} \tau_a$ and $A_\mu =A^a_{\mu } \tau_a$, 
we have
\[
F^a_{\mu \nu} =\partial_\mu A^a_\nu -\partial_\nu A^a_\mu 
-\frac{1}{2} \epsilon^a_{\ bc} (A^b_\mu A^c_\nu -A^b_\nu A^c_\mu )
\]
\begin{equation}
=\partial_\mu A^a_\nu-\partial_\nu A^a_\mu +\epsilon^a_{\ bc}A^b_\mu A^c_\nu .
\end{equation}
Under an infinitesimal gauge transformation, each of the components of $\psi$
 change by $\delta \psi^A$ ( $A=1,2$ ). 
\begin{equation}
\delta \psi^A =\psi^{'A} -\psi^A = \epsilon^a(\tau_a)^A_B\psi^B
\end{equation}
where $\epsilon^a$ ( a=1,2,3 ) are the infinitesimal parameters of 
the gauge transformations. It can be easily shown that under
 these transformations $A^a_\mu$ and $F^a_{\mu\nu}$ change 
 by
 \begin{equation}
 \delta A^a_\mu = \epsilon^a_{\ bc} \epsilon^b A^c_\mu + \partial_\mu \epsilon^a,
\end{equation}
 and 
 \begin{equation}
 \delta F^a_{\mu\nu} = \epsilon^a_{\ bc}\epsilon^b F^c_{\mu \nu}.
 \end{equation}
 Note that $F^a_{\mu \nu}$ transforms as the adjoint representation 
 of the group $SU(2)$. 
 
A \emph{metric} can be defined on the group manifold 
 of $SU(2)$ according to 
 \begin{equation}
 g_{ab}=g_{ba}=\epsilon^c_{\ ad}\epsilon^d_{\ bc}
 \end{equation}
 which is an example of the \emph{Cartan-Killing metric}. 
Note that the isospin indices $a,b, $ are raised and lowered by 
 $g^{ab}$ and $g_{ab}$ in the same manner that the spacetime indices 
 $\mu, \nu,...$ are raised and lowered by the Minkowski metric $\eta _{\mu\nu}$
 or $\eta^{\mu \nu}$. 

The free-field Lagrangian
 density of the Yang-Mills field is 
 \begin{equation}
 {\mathcal L}_{free} =\frac{1}{4} F^a_{\mu\nu} F^{\mu \nu}_a
 \end{equation}
 \\ \ \\
 {\bf \Large 6. Self-duality, instantons, and monopoles}\\

Solutions of Yang-Mills equations which have the important property
 \begin{equation}
 F=\pm \ ^*F
 \end{equation}
are called \emph{self-dual} (+) or \emph{anti-self-dual} (-).
In such a case, the equation of motion $D^*F=0$ reduces
to the Bianchi identity $DF=0$. (Anti)self-dual fields are
therefore solutions of the field equations. 
Among the most 
 interesting solutions of the Yang-Mills equations are instantons and 
 monopoles. The instanton solution of 'tHooft and Polyakov is given by
 \begin{equation}
A_\mu =-\frac{1}{g_o} \frac{\eta_{c\mu\nu}\tau^cx^\nu}{r^2+r_o^2}
 \end{equation}
 where $r_o$ is a constant which prevents $A$ to become singular at $r=0$.
Also $r^2=\sum_{\mu=1}^4 (x^\mu)^2$, $\eta_{abc}=\epsilon_{abc}$ 
 for $a,b,c=1,2,3$, $\eta_{a4b}=-\delta_{ab}$, $\eta_{ab4}=\delta_{ab}$
, and $\delta_{a44}=0$. At large distances, the solution (80) approaches 
the asymptotic form 
\begin{equation}
A_\mu \rightarrow -\frac{i}{g_o}g^{-1}\partial_\mu g
\end{equation}
where $g=(r^2+\lambda^2)^{-1}(x^4+i\tau_a x^a )$. This asymptotic 
form is a pure gauge for which the curvature 2-form vanishes:
\begin{equation}
F=DA=dA+A\wedge A =0
\end{equation}
In other words, the potential 1-form is asymptotically
of the Maurer-Cartan type. The corresponding gauge function $g(x^\mu )$ is 
a mapping from the base space ( $R^4$ ) into the bundle space  or the group 
space. The group manifold of $SU(2)$ is a 3-sphere. As 
 we shall see later, this has important topological implications. 
 
The right action of a group element $h\in G $ on 
 the principal bundle is given by
\begin{equation}
ph=(x,g)h=(x,gh)=p'
\end{equation}
Note that $\pi (p') =\pi (p) =x$. The vector field associated with the 
infinitesimal action 
of the Lie group on the principal bundle is called the \emph{ fundamental
vector field} 
\begin{equation}
\hat{\eta }(p)=\frac{d}{dt}(pe^{t\eta } )
\end{equation}
or
\begin{equation}
\hat{\eta} (x,g) =(x,g\eta )
\end{equation}
Matter fields may couple to the Yang-Mills field. For example, the 
V-valued field $\phi$ transforms as 
\begin{equation}
\phi '=g^{-1} \phi
\end{equation}
 where $g\in G$. In other words, $\phi$ is a mapping from the 
 principal bundle $P$ to the vector space $V$:
 \begin{equation}
 \phi : P\rightarrow V, \ \ \ \ \phi (pg)=g^{-1} \phi (p)
 \end{equation}
 $V$ can be the adjoint representation of $G$. The covariant derivative
 of $\phi $ is given by
 \begin{equation}
 D\phi =d\phi +e[A,\phi ]
 \end{equation}
 The covariant derivative  satisfies the Jacobi relation 
 \begin{equation}
 [D_\mu , [D_\nu , D_\lambda ]] +[D_\nu , [D_\lambda , D_\mu ]]
 +[D_\lambda , [D_\mu , D_\nu ]] =0
 \end{equation}
 The minimal coupling of the $\phi$-field to the Yang-Mills field 
is implemented in the following Lagrangian density
 \begin{equation}
{\mathcal L} =-\frac{1}{2} F\wedge^*F +\frac{1}{2} |D\phi|^2 -U(\phi )
 \end{equation}
 where $U(\phi )$ is the potential term for the $\phi$-field. 
 The field equation for $\phi$ reads
 \begin{equation}
 \ ^*D^*D\phi =-\frac{\partial U}{\partial \phi}, \ \ \ or \ \ \ 
 D_\mu D^\mu \phi^a =-\frac{\partial U}{\partial \phi_a}
 \end{equation}
In terms of components, equation (75) reads
\begin{equation}
D_\mu F^{\mu\nu}=J^\nu.
\end{equation}
 The $\phi$-field provides a current density ( source ) for the 
 Yang-Mills equation 
 \begin{equation}
 J^a_\mu =e[\phi , D_\mu \tau^a \phi ]
 \end{equation}
Note that the underlying symmetry of the Lagrangian demands that 
$J_\mu$ is covariantly conserved ( $D_\mu J^\mu =0$ ),
and at the same time the ordinary conservation law  
\[
 \partial_\mu J^\mu =\partial_\mu \left[ \partial_\nu F^{\nu \mu} + 
 e[ A_\nu , F^{\nu\mu} ]) + e\partial_\mu ([A_\nu , F^{\mu \nu} ] ) 
\right] 
\]
\begin{equation}
=\partial_\mu\partial_\nu F^{\mu\nu } +e \partial_\mu ([ 
 A_\nu , F^{\mu \nu }-F^{\nu \mu } ] ) =0 
 \end{equation}
 is also satisfied. The choice 
 \begin{equation}
 U =\frac{\lambda}{4} (|\phi |^2 -\phi_o^2 )^2 
 \end{equation}
 where $\lambda$ and $\phi_o$ are positive constants leads to the 
\emph{spontaneous breaking} of the gauge symmetry ( \emph{Higgs mechanism} ). This 
 mechanism  is responsible for the formation of massive vector bosons  ( \
 $m_A^2 =e^2 \phi_o^2$ ) . 

 For a static configuration of the Yang-Mills and Higgs fields, the total energy
 functional becomes 
 \begin{equation}
 E=\int \left[ \frac{1}{4} Tr (F_{ij}F^{ij})+ \frac{1}{2} (D_i \phi
 , D^i \phi ) + U (\phi ) \right] d^3x 
 \end{equation}
 In 1974, 'tHooft and Polyakov introduced a static, non-singular solution 
 of the Yang-Mills-Higgs equations with remarkable properties. Consider  
 the following \emph{ hedgehog} ansatz:
 \begin{equation}
\phi^a=G(r) \frac{x^a}{er^2}, \ \ \ and \ \ \  
A^a_i =[ F(r) -1 ] \epsilon_{aij} \frac{x^j}{er^2}, 
\end{equation}
where $F(r)$ and $G(r)$ are unknown functions to be determined by 
 field equations (96) and (97) and $A_o^a=0$. These field equations lead to the following 
 coupled nonlinear differential equations
 \begin{equation}
 r^2 \frac{d^2 F}{dr^2} =FG^2 +F(F^2 -1)
 \end{equation}
 \begin{equation}
 r^2\frac{d^2G}{dr^2} =2F^2 G + \lambda e^{-2} 
 G(G^2 -\mu^2e^2r^2 )
 \end{equation}
These equations can also be obtained
by applying the variational principle to the energy 
functional (94). 
 
 In order to be single valued at $r=0$, $A^a_i$ and 
 $\phi^a$  should  vanish at the origin. This demands 
 \begin{equation}
 F(r)\rightarrow 1, \ \ \ and \ \ \ G(r)\rightarrow 0 \ \ \ as \ \ \ r\rightarrow  0.
 \end{equation}
 For a localized, finite-energy solution, the  gauge field  must
reduce to a pure gauge ( i.e. Maurer-Cartan form ) at large $r$. The 
 $\phi$-field must assume its vacuum ( i.e. $|\phi|\rightarrow \phi_o$ ) 
 far from the origin. Therefore, 
 \begin{equation}
 F(r)\rightarrow 0, \ \ \ and\  \  \ G(r) \rightarrow \mu er \ \ \ 
 as \ \ \ r\rightarrow \infty
 \end{equation}
This behavior guarantees the vanishing of energy density in (94) as 
$r\rightarrow \infty$, since  in this limit,  $F_{ij}\rightarrow 0$, 
$D\phi \rightarrow 0$, and $U\rightarrow 0$. 
In other words, although each component of $\phi$ depends 
on $x^i$, the $\phi$-field is covariantly  "constant". 

 Note that the vacuum of the $\phi$-field corresponds to $U(\phi )=0$. 
 This defines a 2-sphere in the $\phi_a$-space
 \begin{equation}
 \phi_a\phi_a =(\phi_1)^2+(\phi_2)^2 +(\phi_3)^3 =\phi_o^2, 
 \end{equation}
 which is an example of the \emph{ manifold of degenerate vacuua}. 
 The asymptotic form of $F^a_{ij}$ at large $r$ corresponds to a 
 radial, inverse square magnetic field
 \begin{equation}
 B^i =Tr (\epsilon^{ijk} F_{jk}\phi ) =-\frac{x^i}{3r^3}
 \end{equation}
 The full solution to equations (96) and (97) can only be obtained numerically.
 Note that the $SU(2)$ gauge symmetry is 
 spontaneously broken to $U(1)$ at large distances from the monopole 
 ( so is the corresponding bundle ). The Higgs  mechanism leads to the 
 formation of two massive vector fields from the original $A^a_\mu$'s. 
 The remaining massless gauge field is interpreted as the ordinary 
 electromagnetic field  which leads to the identification of the 
 magnetic field (101). 
According to 't Hooft, the following gauge invariant 
quantity properly describes the electromagnetic field
\begin{equation}
F_{\mu\nu} = \hat{\phi}_a F_{a\mu\nu} =
-\frac{1}{g}\epsilon_{abc}\hat{\phi}_aD_\mu \hat{\phi}_b D_\nu \hat{\phi}_c
\end{equation}
where $\hat{\phi}_a =\phi_a/|\phi |$.  With this identification we have
\begin{equation}
\ ^*F^{\mu\nu} =\frac{1}{2g} \epsilon^{\mu \nu \alpha\beta} \hat{\phi}_a 
\partial_\alpha \hat{\phi}_b \partial_\beta \hat{\phi}_c
\end{equation}
and 
\begin{equation}
\partial_\nu\ ^* F^{\mu\nu} =\frac{4\pi}{g} K^\mu
\end{equation}
where $K^\mu$ is the magnetic current. The total magnetic charge is given 
by 
\begin{equation}
Q_m=\frac{1}{g} \int K^0 d^3x =\frac{n}{g}
\end{equation}
where $n$ is the topological charge or winding number. 

There are exact solutions to the field equations in the $\lambda \rightarrow
 0$ limit which is known as the \emph{ Bogomol'nyi-Prasad-Sommerfield}
  ( BPS ) limit (  Prasad and Sommerfield, 1975, and  Bogomol'nyi, 1976 ). 
  It can be shown that a self-dual or anti-self-dual field satisfies the 
Yang-Mills equations. The ( anti ) self-duality 
condition ( $F=\pm \ ^*F$ ) leads to the following first-order equations
\begin{equation}
F=\pm^*D\phi \ \ \ or \ \ \ F_{ij} =\pm \epsilon_{ijk} D_k \phi  .
\end{equation}
These equations are known as the \emph{ Bogomol'nyi equations }. 
For the hedgehog ansatz (95), equations (106) become the following 
first order differential equations for $F(r)$ and $G(r)$
\begin{equation}
r\frac{dG}{dr}=G-(F^2-1),
\end{equation}
\begin{equation}
r\frac{dF}{dr} =-FG.
\end{equation}
These equations can be solved exactly. The monopole solution 
satisfying the appropriate boundary conditions reads
\begin{equation}
G(r)=\mu er \coth (\mu er ) -1 \ \ \ and \ \ \ F(r) =\frac{\mu er}{\sinh (\mu
er)} .
\end{equation}
The reader can easily verify that these solutions satisfy the equations 
(105) and (106) by using the change of variables $1+G=\mu er \psi$ and $F=\mu er 
\chi$.  The Bogomol'nyi equations minimize the energy functional (94), 
with the following total energy 
\begin{equation}
E_{min}=\frac{4\pi \mu |g|}{e}.
\end{equation}

The (anti)self-dual field configurations  are particularly 
important, since they provide the stationary configurations around which
quantum fluctuations can be computed. 
An important property of these models is that the actions are minimized 
at values which are proportional to the corresponding topological 
charges.

If instead of being zero, $A_{a0}=\frac{1}{g}J(r)\frac{x^a}{r^2}$ is 
assumed, we arrive at the following differential equations
\begin{equation}
r^2\frac{d^2K}{dr^2}=K(K^2-J^2+H^2),
\end{equation}
\begin{equation}
r^2\frac{d^2H}{dr^2}=2HK^2 + \lambda H \left[ \frac{1}{g^2} H^2
-r^2F^2 \right],
\end{equation}
and
\begin{equation}
r^2 \frac{d^2J}{dr^2}=2JK^2,
\end{equation}
in which the same ansatz for $A_{ai}$ and $\phi_a$ have been 
used as before. Solutions to these equations are called \emph{ dyons} 
and they bear both electrical and magnetic charges. The electric charge 
of a dyon is given by 
\begin{equation}
Q_e=-\frac{8\pi}{g} \int \frac{JK^2}{r}dr
\end{equation}
and is not necessarily quantized. It can be shown that 
in the BPS limit,
$Q_e=\frac{4\pi}{g}
\sinh \gamma$ where $\gamma$ is an arbitrary constant. 
\\ \ \\
{\bf \Large 7. Topological currents}\\

Perhaps the simplest example of a topological current is the one 
associated with the sine-Gordon system. Consider the sine-Gordon 
Lagrangian density in 1+1 dimensions
\begin{equation}
{\mathcal L}_{SG} =\frac{1}{2} \partial^\mu \phi \partial_\mu \phi -
(1-cos \phi )
\end{equation}
where $\phi$ is a real scalar field on a two dimensional spacetime 
$x^o=t$ and $x^1=x$ with the metric $\eta_{\mu\nu}=diag(-1,1)$. 
The self-interaction potential $V(\phi ) =1-
\cos \phi $ has an infinite number of degenerate vacuua at 
$\phi_n =2n\pi$, $n\in Z$. The Lagrangian density (115) leads 
to the famous sine-Gordon equation
\begin{equation}
\partial^\mu \partial_\mu \phi =sin\phi
\end{equation}
which is known to be an integrable equation with a hierarchy of 
multi-soliton solutions ( Lamb, 1980, and Riazi and Gharaati, 1998 ). Localized, 
finite-energy solutions of (116) satisfy the following boundary 
conditions
\begin{equation}
\phi ( +\infty ) =2n\pi, \ \ \ \phi (-\infty) =2m\pi ,
\end{equation}
where $m$ and $n$ are integers. Topological current for the sine-Gordon
system is given by
\begin{equation}
J^\mu =\frac{1}{2\pi} \epsilon^{\mu\nu} \partial_\nu \phi,
\end{equation}
where $\epsilon^{\mu\nu}$ is the totally antisymmetric tensor in two 
dimensions. The current density  (116) is conserved
\begin{equation}
\partial_\mu J^\mu =\frac{1}{2\pi} \epsilon^{\mu \nu} 
\partial_\mu \partial_\nu \phi =0
\end{equation}
since $\epsilon^{\mu\nu}$ is antisymmetric while 
$\partial_\mu \partial_\nu$ is symmetric. 
The total charge of a localized solution ( both static and  
dynamic ) is easily shown to 
be quantized
\[
Q_{SG}=\int_{-\infty}^{+\infty} J^o dx =
\frac{1}{2\pi} \int_{-\infty}^{+\infty} \epsilon ^{o1}\partial_1 \phi dx
\]
\begin{equation}
=\frac{1}{2\pi} \int_{-\infty}^{+\infty} \frac{\partial \phi}{\partial
x } dx \\ =\frac{1}{2\pi}\left[ \phi(+\infty ) - \phi (-\infty ) 
\right] =n-m
\end{equation}
in which the boundary conditions (117) are used. Note that solutions with 
different topological charges belong to distinct topological sectors. 
They are separated from each other by infinite energy barriers. In other
words, they cannot be continuously deformed into each other. 

The concept of topological
charges can be extended to more complicated fields in higher dimensions. 
Consider, for example, a time-dependent complex scalar field $\phi =
\phi_1 + i\phi_2    $ on the complex plane $z$. The vacuum manifold of 
$\phi$ is assumed to reside at
\begin{equation}
|\phi|^2 =\phi_1^2+\phi_2^2 =\phi_o^2
\end{equation}
where $\phi_o$ is a real, positive constant. The topological current for 
this field can be defined  as
\begin{equation}
J^\mu = \frac{1}{2\pi \phi_o^2} \epsilon^{\mu \nu \alpha } \epsilon_{ab} \partial_\nu 
\phi_a \partial_\alpha \phi_b
\end{equation}
where $\mu,\nu,\alpha =0,1,2$ and $a,b=1,2$. It can be easily shown 
that this  current is conserved ( $\partial_\mu J^\mu =0$ ). 
The total charge of a localized field configuration on 
the complex plane ( $z=x+iy$ ) is 
\begin{equation}
Q=\int\int J^o dxdy = \frac{1}{2\pi \phi_o^2}\int \int \epsilon^{ij} \epsilon_{ab}
\partial_i \phi_a \partial_j \phi_b dxdy .
\end{equation}
This can be rewritten as $\int \int \vec{\nabla}\times
\vec{H} .d\vec{S}$ where $d\vec{S} =dxdy \hat{k}$, and 
$\vec{\nabla}\times \vec{H}.\hat{k} =\epsilon_{3ij}
\partial_i H_j$ with 
\begin{equation}
H_j=\frac{ \epsilon_{ab} \phi_a \partial_j\phi_b }{2\pi\phi_o^2} 
\end{equation}
Note that a $z$-axis is artificially introduced to facilitate the 
notations of ordinary vector analysis. $\hat{k}$ is the unit vector in the positive 
$z$-direction. Stokes's theorem can now be used
\begin{equation}
Q=\oint_C \vec{H}.d\vec{l}
\end{equation}
where $d\vec{l}$ is a line element along the boundary curve $C$ which is 
assumed to be in the form of a circle on the complex plane with radius
$r\rightarrow \infty $. Parametrizing $dl$ and $d\phi $ along $C$ according 
to 
\begin{equation}
dl=rd\theta, \ \ \ \ d\phi=\phi_o d\alpha ,
\end{equation}
leads to 
\[
Q=\frac{1}{2\pi \phi_o^2} \int^{2\pi}_{\theta =0} \frac{1}{2}
(\phi_1 \frac{\partial \phi_2}{\partial \theta } 
-\phi_2 \frac{\partial \phi_1}{\partial \theta } ) 
rd\theta 
\]
\begin{equation}
=\int^{2\pi }_{\theta =0} \frac{d\alpha }{d\theta } 
d\theta =\frac{1}{2\pi } \oint^{2\pi}_{\theta =0} d \alpha (\theta )=n
\end{equation}
where $n$ is an integer corresponding to the number of times the $\phi$ 
field circles its vacuum $S^1$ as the path $C$ is completed \emph{ once} 
on the complex plane. We saw in section 3 that $n$ was properly called the 
winding number. 

For an isovector field $\phi_a$ ( $a=1,2,3$ ) 
with an $S^2$ vacuum
\begin{equation}
\phi_a\phi_a=\phi_o^2
\end{equation}
the topological current can be defined as ( Vasheghani and Riazi, 1996 )
\begin{equation}
J^\mu =\frac{1}{4\pi \phi_o^3} \epsilon^{\mu \nu \alpha \beta } 
\epsilon_{abc} \partial_\nu \phi_a 
\partial_\alpha \phi_b \partial_\beta \phi_c
\end{equation}
Note that the spacetime is now the ordinary Minkowski spacetime and 
$\mu ,\nu,... =0,1,2,3$ with $x^o= t$. Once again, the current is 
identically conserved ( $\partial_\mu J^\mu =0$ ), and the total 
charge is quantized
\begin{equation}
Q=\int J^o d^3x =\frac{1}{4\pi \phi_o^3} \oint \frac{dS_\phi}{dS_x} dS_x
=n
\end{equation}
where $dS_\phi$ and $dS_x$ are area elements in the $x$-space and 
$\phi$-space, respectively. The current (129) can be written as the 
covariant divergence of an anti-symmetric second-rank tensor 
\begin{equation}
\partial_\mu F^{\mu \nu } =J^\nu
\end{equation}
where
\begin{equation}
F^{\mu \nu} =\frac{1}{4\pi \phi_o^3 } \epsilon^{\mu \nu \alpha \beta}
\left[ \epsilon_{abc} \phi_a\partial_\alpha \phi_b \partial_\beta \phi_c
+\partial_\beta {\mathcal B}_\alpha \right]
\end{equation}
in which ${\mathcal B}_\alpha $ is an auxiliary vector field. It is 
interesting to note that the dual field $\ ^*F$ with the following 
tensorial components 
\begin{equation}
\ ^*F^{\mu \nu} =\frac{1}{2} \epsilon^{\mu \nu \alpha \beta } F_{\alpha
\beta } 
=2\epsilon_{abc} \phi_a \partial^\mu \phi_b \partial^\nu \phi_c 
+\partial^\mu {\mathcal B}^\nu -\partial^\nu {\mathcal B}^\mu 
\end{equation}
satisfies the equation
\begin{equation}
\partial_\mu \ ^*F^{\mu \nu } = 0
\end{equation}
provided that the vector field ${\mathcal B}^\mu$ is a solution of the 
following wave equation
\begin{equation}
\Box {\mathcal B}^\mu  -\partial^\mu (\partial_\alpha {\mathcal B}^\alpha )
=2\epsilon_{abc} \partial _\alpha ( \phi_a \partial^\mu 
\phi_b \partial^\alpha \phi_c  ) 
\end{equation}
The right hand side of this equation defines another conserved current 
\begin{equation}
K^\mu = 2\epsilon_{abc}\partial_\alpha ( \phi_a \partial^\mu \phi_b
\partial^\alpha \phi_c )
\end{equation}
which is consistent with the vanishing of the divergence of the left
hand side of equation (135).
The resemblance of equations (131) and (134) to Maxwell's equations and the 
capability of this model to provide non-singular models of charged particles
is discussed in Vasheghani and Riazi ( 1996 ). 
Let us write (132) in the following form
\begin{equation}
G=F-H
\end{equation}
where 
\begin{equation}
G^{\mu \nu} =\frac{1}{4\pi \phi_o^3} \epsilon^{\alpha\beta\mu\nu} 
\epsilon_{abc} \phi_a \partial_\alpha \phi_b \partial_\beta \phi_c 
\end{equation}
and 
\begin{equation}
H^{\mu\nu} =\frac{1}{4\pi \phi_o^3} \epsilon^{\mu\nu\alpha\beta} 
\partial_\beta \mathcal{B}_\alpha 
\end{equation}
We now have 
\[
dF=0,
\] 
and 
\begin{equation}
d^*H =0. 
\end{equation}
Any 2-form like $G$ which can be 
written as the sum of two parts satisfying (140) is said to admit 
\emph{ Hodge decomposition }. We shall see in the next section that 
forms like $G$ and $H$ are cohomologous ( i.e. they belong to the 
same cohomology class ), since they differ only by an exact form. 

Topological currents of defects in various space dimensions can 
be formulated in a unified way. 
Duan et al. ( 1999 ) considered the following topological current for 
point defects in a medium represented by an $n-$dimensional 
order parameter $\phi^a$ ( $a=1,...,n$ ):
\begin{equation}
J^\mu =\frac{1}{A(S^{n-1})(n-1)!}\epsilon^{\mu \mu_1 ...\mu_n}
\epsilon_{\hat{\phi}_1 ... \hat{\phi}_n} \partial_{\mu_1}\hat{\phi}^{a_1}...\partial_{\mu_n}
\hat{\phi}^{a_n}
\end{equation}
where $A(S^{n-1})=\frac{2\pi^{n/2}}{\Gamma (n/2)}$ is the area 
of the $(n-1)$-dimensional unit sphere, and 
\begin{equation}
\hat{\phi}^a=\frac{\phi^a}{|\phi |}.
\end{equation}
The number of space dimensions is also assumed to be $n$.
The topological charge density $\rho =J^o$ which represents the defect density,
is everywhere zero except at the location of the defects where it diverges.
It therefore behaves like a delta function, and can be represented as 
\begin{equation}
J^\mu =\delta (\phi )D^\mu (\frac{\phi }{x} )
\end{equation}
where $\delta (\phi )$ is the Dirac delta function, and the 
$n-$dimensional Jacobian determinant is defined by
\begin{equation}
\epsilon^{a_1...a_n}D^\mu (\frac{\phi}{x})=
\epsilon^{\mu \mu_1 ...\mu_n}\partial_{\mu_1} \phi^{a_1}...\partial_{\mu_n}
\phi^{a_n}.
\end{equation}
The defects are therefore located at points where the equations
\begin{equation}
\phi^a =0
\end{equation}
are satisfied. The defect density is given by  ( Liu and Mazenko, 1997 ):
\begin{equation}
\rho =J^o=\delta (\phi )D^o (\frac{\phi}{x} ) 
\end{equation}
where $D^o=|\frac{\partial (\phi^1 ...\phi^n )}{\partial (x^1,...,x^n)}|$
 is the ordinary Jacobian determinant.  The total topological 
charge is given by
\begin{equation}
Q=\int J^o d^n x =\sum_{i=1}^l \beta_i \eta_i
\end{equation}
where $\beta_i$ and $\eta_i$ are the Hopf indices and Brouwer 
degrees of the $\phi$ mapping, respectively, and $l$ is the number 
of point defects in the system. 
\\ \ \\
{\bf \Large 8. Cohomology and electromagnetism} \\

The theory of cohomology groups was developed by G. de Rham in 1930's. 
In this section we present a brief introduction to  the subject. 
The relevance of cohomological methods to other gauge fields 
is discussed in ( Henneaux, 1988 ).

Closed $p-$forms on a manifold ${\mathcal M}$ form a  vector space denoted by 
$C^p ({\mathcal M} )$. These forms are also called \emph{p-cocycles}. The subspace 
of $C^p ({\mathcal M} )$  which comprises exact $p-$forms is denoted 
by $B^p ({\mathcal M})$. Exact and closed $p-$forms are called \emph{ 
p-coboundaries }. The $p$-th cohomology group is defined as the 
quotient group $C^p({\mathcal M})/B^p( {\mathcal M})$:
\begin{equation}
H^p ( {\mathcal M} ) =C^p ( {\mathcal M} ) / B^p ( {\mathcal M} )
\end{equation}
The elements of $H^p( {\mathcal M} )$ form equivalence classes which are 
called \emph{p-th cohomology classes}.

Two closed $p-$forms $\omega$ and $\eta$ are \emph{ cohomologous } 
( i.e. belong to the same cohomology  class ) if their difference 
is an exact form 
\begin{equation}
\omega -\eta =d \rho 
\end{equation}
where $\rho$ is a $(p-1)-$form. The $p$-th de Rham cohomology group is the 
set of these equivalence classes. Note that the 0-cohomology class is the 
class of exact $p-$forms ( in other words, all exact $p-$forms are 
cohomologous ). 

Let us list a few well-known results  relevant to the cohomology groups 
( Guillemin and Pollack, 1974 )
\begin{itemize}
\item{ Exact 0-forms do not exist.}
\item{$H^p (R\times M )$ is isomorphic to $H^p(M)$.}
\item{$H^p(R^k) =0$ if $k> 0$ and $p> 0$. Every closed 
$p-$form ( $p> 0$ ) on $R^k$ ( $k>0$ ) is exact.}
\item{The previous item can be generalized to 
$H^p({\mathcal M})=0$  for all $p>0$ if ${\mathcal M}$ is contractible.}
\item{$H^p (S^k )$ is one dimensional for $p=0$ and $p=k$. For all
other $k>0$, $H^p ( S^k ) =0$.}
\item{$H^o ({\mathcal M})$ is the space of constant functions on ${\mathcal M}$
and its dimension counts the number of connected pieces of the manifold. 
( e.g. $dim H^o ( R^n ) =1$ ). Manifolds which have  globally trivial 
coordinates ( $e.g.\ \  R^n$ ) have trivial de Rham cohomologies 
( $H^p ( R^n ) =0$ for $p>0$ ). }
\item{The \emph{Euler-Poincare' characteristic} 
of the manifold ${\mathcal M}$ is defined by 
\begin{equation}
\chi_{\mathcal M} =\sum_{p=0}^n (-1)^p d_p
\end{equation}
where $d_p =dim H^p( {\mathcal M} )$ is the dimension of the $p-$th 
homology group ( called the \emph{ Betti number } ).  For an 
$n-$dimensional sphere $S^n$, $H^o (S^n ) =H^n (S^n ) = R$ 
and $H^p (S^n )=0$ for $0< p < n$. We therefore have $\chi ( S^n ) =0$ 
for $n=odd$ and $\chi ( S^n ) =2  $ for $n=even$. 
For odd-dimensional manifolds, the Euler characteristic vanishes as we saw 
in the above  example.} 
\end{itemize}

In the absence of magnetic 
monopoles $dF=0$ where $F$ is the electromagnetic field 2-form. $F$ is 
therefore a closed 2-form. 
It is also exact and can be written as $F=dA$ where $A$ is the 
electromagnetic potential 1-form. The phase
factor $exp(i\oint_c A )$ which appears in the quantum mechanical context  
( e.g.  the  Aharonov-Bohm effect ) 
can be written as 
\begin{equation}
exp(i\oint_C A )=exp(i\int_S F )
\end{equation}
using the Stokes's theorem. Here, $S$ is a surface
area with the closed boundary curve $C$. But $\int_S F$ 
is the magnetic flux passing through $S$ . The phase factor (151) is known to be observable as shifts in 
in the interference fringes of electrons in the  Aharonov-Bohm 
experiment. This shows that  although $A$ is ambiguous 
up to a $U(1)$ gauge transformation, it cannot be assumed redundant at a
quantum level. In the presence of magnetic monopoles, 
$F$ is no longer closed, and cannot 
be derived from a 1-form. A single-valued 
gauge  transformation in the  region surrounding a magnetic  monopole
( over which $F$ is closed ), leads to a Dirac relation between the electric 
and magnetic charges ( Wu and Yang, 1975 )
\begin{equation}
g=\frac{1}{e} 
\end{equation}

In section 4 we saw that the field of a magnetic monopole
can be obtained from the potential 1-form 
\begin{equation}
A=-\frac{g}{4\pi} cos \theta d\phi. 
\end{equation}
Although this potential looks smooth over a 2-sphere, the coordinate 
system itself is singular along the z-axis.  This gauge potential, therefore,
is not defined on the z-axis. When transformed into Cartesian coordinates, 
this potential becomes 
\[
-\frac{g}{4\pi} \left( 
-\frac{zy}{r(x^2+y^2)}, +\frac{zx}{r(x^2+y^2)}, 0\right) 
\] 
which clearly shows 
the singularity along the z-axis. 

The gauge potential of a magnetic monopole can be  chosen as
\begin{equation}
A_\pm =\frac{n}{2r} \frac{xdy-ydx}{z\pm r}
\end{equation}
where $A_\pm $ are the potentials over the northern (+) and 
southern (-) hemispheres of an $S^2$ centered on the monopole. Over 
 an equatorial strip, $A_\pm$ are related 
 by $A_+ =A_- +nd\phi$ which shows that $A_+$ and $A_-$ do not merge 
 smoothly unless $n=0$. Note that $\oint_{S^2}F=2n\pi $ and therefore $n$ 
 represents the quantized magnetic charge of the monopole. The 
 monopole singularity at $r=0$ should be considered  as a hole 
 in the base manifold. The magnetic charge  arises from the non-trivial
topology of the principal bundle. Occurrence of such integers 
associated with  non-trivial 
bundles are properly described by the concept of characteristic classes, to 
be discussed briefly in the next section.

Electric and magnetic charges act as holes in the base manifold 
( Minkowski spacetime in this case ). Outside these holes, $F$ and $\ ^*F$ 
are closed and their integral over
a closed surface are integer multiples 
of elementary electric and magnetic charges. 
In fact, even in the absence of magnetic and electric 
charges, topologically non-trivial curved spaces can 
lead to similar effects. Consider, for example, the source-free 
Maxwell's equations on the following so-called \emph{ wormhole} spacetime
\begin{equation}
d\tau^2 =dt^2 -dr^2 -(r^2+r_o^2) 
( d\theta^2 + \sin^2 \theta d\phi^2 )
\end{equation}
in which $r_o > 0$ is called the throat radius ( Morris and Thorne, 1988 ).
Note that $r$ extends from $-\infty$ to $+\infty$, and the metric (155) 
describes two asymptotically Minkowskian spacetimes joined by an $S^2$. 
Covariant Maxwell's equations $F^{\mu\nu}_{;\mu} =0 $ and 
$\ ^*F^{\mu\nu}_{;\mu}=0$ have the following non-singular solutions
\begin{equation}
E(r)=\frac{Q_E}{r^2+r_o^2}; \ \ \ \ B(r)=\frac{Q_M}{r^2+r_o^2}
\end{equation}
where $E$ and $B$ are the radial electric and magnetic fields and $Q_E$ and 
$Q_M$ are constants of integration. 
The Stokes's theorem $\int_V dF =\int_{\partial V} F$ 
and a similar relation for $\ ^*F$, when applied to either side of 
the wormhole imply that the electric and magnetic fluxes through any
2-sphere centered at the wormhole are independent of $r$. Here,
$V$ is a 3-volume confined between two parallel 2-spheres. 
\\ \ \\ {\bf \Large 9. Homotopy groups and cosmic strings }\\

Let $C_1$ and $C_2$ be two loops in a manifold 
${\mathcal M}$, based at a point $x_o$. These two loops are
\emph{homotopic} (denoted by 
$C_1\simeq C_2 $) if they can be continuously deformed into each other. 
This corresponds to the existence of a continuous set of curves $C(\lambda )$
with $\lambda \in [0,1]$ such that $C(0)=C_1$ and $C(1)=C_2$.
Homotopy is an equivalence relation and the set of all closed 
curves in ${\mathcal M}$ are divided into a set of homotopy classes. The composite
curve $C_3=C_2 \circ C_1$ is defined according to 
\begin{equation}
C_3( \lambda )=
\left\{ \begin{array}{ll} \mbox{$C_2(2\lambda )$} & \mbox{$ if \ \ \ 0\leq \lambda \leq 1/2,$} 
\\
\mbox{$C_1 ( 2\lambda -1 ),$} & \mbox{$if \ \ \ 1/2\leq \lambda \leq 1.$}
\end{array} \right.
\end{equation}
Homotopic deformations of any class $C$ provide the identity ($Id$) of
that class.  The inverse of 
a homotopy $C^{-1}$  is defined according to the relation 
$C^{-1}\circ C = C\circ C^{-1} =Id$ . A loop is homotopic to 
zero if it can be continuously deformed into a point. Such
loops are also called \emph{ null-homotopic}. 
Homotopy classes with the inverse and identity defined in this way form 
a group called the first or \emph{ fundamental group} denoted by 
$\pi _1$. A manifold is called \emph{ simply connected} if all 
loops in it are null-homotopic. $R^n$ is an obvious example of a simply 
connected manifold ( $\pi_1 (R^n)=0$ ). The group manifold of $U(1)$ 
is a circle ( $S^1$ ). The fundamental group  of $U(1)$ therefore 
corresponds to the number of times a loop circles the group manifold in 
the clockwise or counter-clockwise direction. We therefore have
$\pi_1 (S^1 )=Z$, and the same is for $R^2-\{ 0\}$. 
Other examples of fundamental groups include $\pi_1(O(2))=\pi_1 (SO(2)) =
\pi_1 (U(N))=Z$, while $\pi_1 (SO(3))=\pi_1 (SO(N))=\pi_1(O(N)) =Z_2$ 
( $N\geq 3$ ), where $Z_2$ is the group of integers modulo 2. 
Furthermore, $SU(N)$ groups are simply connected and their fundamental 
group vanishes. 

Higher homotopy groups are defined in a similar manner using compact 
hypersurfaces homotopic to $S^n$ ( i.e. continuously deformable to 
an $n-$dimensional sphere ). The $n-$th homotopy group $\pi_n$ therefore 
comprises the homotopy classes of maps from $S^n$ to the 
manifold under consideration. In particular, we have $\pi_n ( S^n)=Z$, 
$\pi_3 ( S^2) =Z$ , and since the 
group manifold of $SU(2)$ is $S^3$, $\pi_3( SU(2)) =Z$. 

A manifold ${\mathcal M}$ is \emph{ p-connected} if all homotopy groups 
$\pi_i ( {\mathcal M} ) $ vanish for $i\leq p$. 

The wormhole space considered in the previous section is 
simply connected but not 2-connected, since 2-spheres which 
contain the wormhole cannot be contracted to a point.
Homotopy groups have important implications for the  existence and
 stability of topological defects and solitons. 
Consider, for example, 
a complex scalar field $\phi$ with the self-interaction potential 
\begin{equation}
V(\phi ) = \frac{\lambda }{4} ( \phi ^* \phi - \phi_o^2 )^2 
\end{equation}
where $\lambda$ and $\phi_o$ are constants.  The vacuum manifold is a 
$S^1$ on the complex $\phi$ plane. 
We can now re-interpret the contents of section 7 ( equation 121 
onwards ) in the 
framework  of homotopy groups. For a localized field on the 
$xy$-plane, the $\phi (x,y)$ field along a large circle $x^2+y^2=r^2$ 
( $r\rightarrow \infty$ ) belongs to a homotopy class mapping the 
circle $S^1: x^2+y^2=r^2$ into the circle $S^1: \phi_1 
^2 +\phi_2^2=\phi_o^2$. This is nothing but the fundamental group $S^1$: 
$\pi_1 (S^1) =Z$. The topological charge ( 125 ) corresponds to the degree
of this mapping and labels the corresponding homotopy class. However, the 
Lagrangian density
\begin{equation}
{\mathcal L} =\partial^\mu \phi^* \partial_\mu \phi -\frac{\lambda}{4} 
(\phi^*\phi -\phi_o^2)^2 
\end{equation}
does not  lead to localized finite energy solutions. 
This problem can be demonstrated by the $Q=1$ sector with the asymptotic 
behavior $\phi =\phi_o \exp ( i\theta)$ in which $\theta = \arctan (y/x)$. 
The presence of the $|\nabla \phi |^2$ term in the Hamiltonian 
density
\begin{equation}
{\mathcal H} = \frac{1}{2}|\nabla \phi |^2+V(\phi )
\end{equation}
leads to an energy density 
\begin{equation}
u_\phi \simeq |\nabla \phi |^2 =\phi_o^2 (\nabla \theta
)^2 =\frac{\phi_o^2}{ r^2}
\end{equation}
in plane polar coordinates. This, however, leads to a logarithmically 
divergent energy integral
\begin{equation}
\int^\infty \frac{\phi_o^2}{r^2}2\pi rdr = \phi_o^2 \ln r|^\infty  \rightarrow \infty
\end{equation}
In order to cure this problem, the global $U(1)$ symmetry of the Lagrangian
(159) can be made local
\begin{equation}
{\mathcal L} = D^\mu \phi^* D_\mu \phi -\frac{\lambda}{4} (\phi^* \phi 
-\phi_o^2 )^2 -\frac{1}{4} F^{\mu\nu} F_{\mu \nu} 
\end{equation}
This leads to the so-called the \emph{ Abelian  Higgs model}.
We demand that the $\phi$ field be covariantly constant at large $r$
\begin{equation}
D\phi =0 \ \ \  \rightarrow \vec{\nabla} \phi -ie\vec{A} =0
\end{equation}
with the asymptotic value $\phi =\phi_o \exp (i\theta )$.
Equation (164) gives
\begin{equation}
\vec{A} =\vec{\nabla } ( -\frac{i\phi}{e} ) \ \ \ as \ \ \ 
r\rightarrow \infty .
\end{equation}
The potential 1-form $A$ is therefore a pure gauge and leads to 
a vanishing curvature $F$ at large distances. The energy density 
of the gauge field, therefore, also vanishes at large distances. 
Such a solution is non-trivial, since it corresponds to a non-vanishing
magnetic flux through the xy-plane
\begin{equation}
\Phi_B =\int F =\int dA =\oint A =\oint \vec{\nabla} ( -
\frac{i\phi}{e}) .\vec{dl} = \frac{1}{e} \phi_o \oint d\theta =
\frac{2\pi}{e} \phi_o
\end{equation}
where the integration is performed over an infinitely large 
disk.
Solutions belonging to other homotopy classes with topological charges 
(127) have magnetic fluxes $\Phi_B = \frac{2\pi n}{e} \phi_o$. These
solutions which are in the form of bundles of magnetic lines of force in a 
three dimensional space are called \emph{ cosmic strings} in
cosmological terminology. They 
can also represent magnetic flux tubes in the Landau-Ginzburg model of 
superconductivity. 

For a cosmic string, we have a smooth map from the 
circle $S^1$ in the configuration space to the $U(1)$ manifold which was
also $S^1$. The Brouwer degree or winding number of this map corresponds
exactly to the integers  which label the first homotopy group 
$\pi_1(S^1)$. In this case, the winding number is given by 
\begin{equation}
n=\frac{1}{2\pi \phi_o^2} <^*d\phi |id\phi >.
\end{equation}
Note that ( Felsager, 1983 )
\[
<^*d\phi |id\phi >=-i\int_{R^2} d\bar{\phi} \wedge d\phi =
-i\int_{R^2}d(\bar{\phi}d\phi ) 
\]
\begin{equation} 
 =  \lim _{r_o\rightarrow \infty} 
-i\int_{\rho =\rho_o} \bar{\phi}d\phi =-i\phi_o^2 
\lim_{\rho_o\rightarrow \infty } \int_{\rho=\rho_o} 
id\phi = 2n\pi \phi_o^2
\end{equation}
in which $\bar{\phi}$ is the complex conjugate of $\phi$.

The model described above can also be applied to the magnetic 
flux tubes which form in  superconductors.
Electrons in a superconductor form pairs. These so-called \emph{ Cooper pairs} can 
be described by a complex scalar field $\phi$ which is called the \emph{ 
order parameter}.  $|\phi|^2$ represents the density of the Cooper pairs. 
The energy density of a static configuration is given by 
\begin{equation}
{\mathcal H}= \frac{1}{2} |\vec{\nabla}\phi |^2 +\frac{1}{2} 
\alpha |\phi |^2 +\frac{1}{4} \beta |\phi|^4 + C
\end{equation}
where $C$ is a constant and $\alpha$ is a parameter which depends on 
the temperature via 
\begin{equation}
\alpha (T) =a\frac{T-T_c}{T_c}
\end{equation}
in which $a$ is 
a positive constant and $T_c$ is the critical temperature. For $T>T_c$, 
the state of minimum energy density happens at $\phi =0$ while for 
$T<T_c$, the minimum converts into a $S^1$ in the $(Re(\phi ), Im(\phi))$ plane
\begin{equation}
|\phi|^2=-\frac{\alpha}{\beta} >0
\end{equation}
The quantity $\xi (T)\equiv \frac{1}{\sqrt{-\alpha }}$ has the dimensions 
of length and is called the \emph{coherence length} of the superconductor. 
In the presence of a magnetic field, we have to implement the covariant 
derivative
\begin{equation}
D=d-i\frac{g}{\hbar }A
\end{equation}
where $A$ is the magnetic vector potential, and include 
the EM energy density in (169). 

It can be shown that $\vec{B}$ 
satisfies
\begin{equation}
\nabla^2\vec{B} +\frac{\alpha q^2}{\beta \hbar^2} \vec{B} =0
\end{equation}
showing that the magnetic field is exponentially damped inside a 
superconductor. The quantity $\sqrt{-\alpha /\beta} q/\hbar $ is called 
the \emph{ penetration length} of the superconductor.
The reader notes that the Abelian Higgs model is mathematically equivalent 
to the Ginzburg-Landau theory of superconductivity. The following 
correspondence can be stablished between the parameters of the two models
\begin{equation}
e\leftrightarrow g/\hbar, \ \ \ \lambda \leftrightarrow \beta, \ \ \ and \ \ \ 
\mu^2 \leftrightarrow -\alpha .
\end{equation}

It can be visualized that a pair of magnetic monopoles with opposite charges
may reside at the 
end-points of a finite string. The quantization of magnetic 
charge is consistent with the quantization of the magnetic flux through 
the string ( $\Phi_B =g=\frac{2\pi n}{e}$ ). Since any increase in the 
distance between the monopole pair is associated with an equal increase
in the length of the string, a linearly increasing potential between the 
monopoles is implied ( Felsager, 1983 )
\begin{equation}
V(r)=\Lambda r
\end{equation}
where $\Lambda$ is the mass per unit length of the string.
This model provides a possible mechanism for the confinement of  magnetic 
monopoles. A similar mechanism  was suggested for the quark confinement 
inside the hadrons ( Nambu, 1985 ).

A rotating relativistic string has the interesting property that its 
angular momentum $J$ is proportional to its mass squared ( $M^2$). This
property is observed  in the \emph{Regge trajectories} of the baryons with the 
same isospin and strangeness. A simpler realization of confinement for 
solitons with fractional topological charges was introduced by 
Riazi and Gharaati ( 1998 ). 
\\ \ \\
{\bf \Large 10. Characteristic classes}\\

In dealing with non-trivial fiber bundles, transition functions 
and integrals of the curvature 2-form lead to integers which have a 
topological origin. Characteristic classes are an efficient and elegant 
way to distinguish and classify inequivalent fiber bundles. 

Consider a $k\times k$ complex matrix $m$. A polynomial $P(m)$ constructed 
with the components of $m$ is a  \emph{ characteristic polynomial} if 
\begin{equation}
P(m)=P(g^{-1}mg) \ \ \ for \ \ \ all\ \ \ g\in GL(k,C)
\end{equation}
where $g$ represents a complex matrix belonging to  general linear transformations.
For example, 
\begin{equation}
\det ( 1+m)=1+S_1(\lambda )+...+S_k(\lambda )
\end{equation}
is an invariant polynomial constructed with the $i-$th symmetrical 
polynomial 
\begin{equation}
S_j (\lambda ) =\sum_{i_1 <...<i_j} \lambda_{i_1}...\lambda_{i_j}
\end{equation}
where $\lambda_i  $s are the  eigenvalues of $m$. Curvature 2-forms $\Omega$ 
which are matrix-valued have $P( \Omega )$ which are closed and have 
invariant integrals ( Chern, 1967 ). The \emph{ total Chern form} 
is defined as
\begin{equation}
C(\Omega ) = \det ( 1+ \frac{i}{2\pi} \Omega ) = 1 + c_1( \Omega ) +
c_2 (\Omega ) +...
\end{equation}
where $c_i (\Omega )$ is a polynomial of degree $i$ in the curvature 
2-form $\Omega$:
\[
c_o =1; \ \ \ \ c_1 =\frac{i}{2\pi } Tr \Omega,
\]
\begin{equation}
c_2 = \frac{1}{8\pi^2} \{ Tr ( \Omega\wedge \Omega ) - Tr\Omega \wedge 
Tr \Omega \}, \ \ \  etc.
\end{equation}
Note that $c_i =0$ for $2i > n$ where $n$ is the dimension of the 
base manifold. Closedness causes the Chern forms $c_i (\Omega )$ belong to 
distinct cohomology classes. These classes have integer coefficients. 
Integrals like 
\begin{equation}
\int_{\mathcal{M}} c_2(\Omega )
\ \ \ \ and \ \ \ \ \int_{{\mathcal M}} c_1(\Omega )\wedge c_1(\Omega )
\end{equation}
are invariant integers called \emph{ Chern numbers }. 

The action for (anti)self-dual field configurations is proportional to 
the second Chern number
\begin{equation}
S=-\frac{1}{2}\int Tr F\wedge^*F =\mp \frac{1}{2} \int TrF\wedge F
=4\pi |C_2|
\end{equation}
where $C_2=\frac{1}{8\pi } \int Tr F\wedge F$ is the second Chern 
number. The $|C_2|=1$ case corresponds to the 't Hooft instanton. 

The Chern form is a global form on the base manifold and does not depend on 
 the frame chosen. Chern classes are also closely related to the homotopy 
 theory, since the set of isomorphic classes of $i-$dimensional vector 
 bundles is isomorphic to the homotopy classes of maps from the base 
 manifold  to $Gr(m,i,C)$ where $Gr(m,i,C)$ is the Grassmann manifold of 
 i-planes in $C^m$. 

 As a simple example, consider the $U(1)$ bundle of the Dirac monopole 
 over  a $S^2$. We have
 \begin{equation}
 det (1+\frac{i}{2\pi } \Omega ) =1+\frac{i}{2\pi } \Omega
 \end{equation}
 and therefore $c_1 =i\Omega /2\pi$. Note that $c_1$ is real since 
 $\Omega $ is pure imaginary ( $\Omega =iF=idA$ ). 
 Furthermore, using equation (154) we have 
 \begin{equation}
 \int_{S^2} c_1 = -\frac{1}{2\pi} 
 \int_{S^2_+}dA_+ -\frac{1}{2\pi} \int_{S^2_-} dA_- =
 -\frac{1}{2\pi } \int_o^{2\pi} nd\phi =-n
 \end{equation}
 which shows that  the Chern number and the monopole charge are essentially the 
 same for this example. Note that the integration over $S^2$ is divided into 
 two hemispheres $S^2_+$ and $S^2_-$ and $dA_+-dA_-=nd\phi$ is used. 

 The characteristic classes of real vector bundles are called \emph{ 
 Pontrjagin classes}. In similarity with the 
definition (179), the \emph{ total Pontrjagin class} of a real $O(k)$ 
 bundle is defined as 
 \begin{equation}
 P(\Omega ) =\det (1-\frac{1}{2\pi} \Omega) = 1+p_1+p_2+...
 \end{equation}
where $\Omega$ is the bundle curvature. The orthogonality conditions 
lead to the vanishing of odd-degree polynomials. The invariant polynomials
are closed here also and  the resulting cohomology classes are independent
of the connection form. 

One can also attribute Pontrjagin classes to the electromagnetic 
field in the following way ( Eguchi et al., 1980 )
\begin{equation}
\det(1-\frac{1}{2\pi} F ) =1 + p_1+p_2
\end{equation}
where $F$ is the electromagnetic field tensor in its matrix form, 
and $p_1$ and $p_2$ are related to the EM energy density and 
Poynting vector according to 
$p_1=\frac{1}{(2\pi)^2} (E^2+B^2)$ and $p_2=\frac{1}{(2\pi)^4} (\vec{E}.
\vec{B})^2 $.
\\   \\
{\bf \Large 11. Differential geometry and Riemannian manifolds}\\

We closely follow Eguchi et al. ( 1980 ) in this section.
The metric on a Riemannian manifold ${\mathcal M}$ can be written in the 
form 
\begin{equation}
ds^2=g_{\mu\nu} dx^\mu dx^\nu = e^a\eta_{ab}e^b
\end{equation}
where $e^a=e^a_\mu dx^\mu $ is the vierbein basis of $T^*({\mathcal M})$ 
and $\eta_{ab}$ is the flat metric ( $\eta_{ab}=\delta_{ab}$ for 
a Euclidean manifold ). The connection 1-form $\omega^a_b =\omega^a_{b\mu}
dx^\mu$ obeys the Cartan structure equations
\begin{equation}
T^a=de^a+\omega^a_b\wedge e^b
\end{equation}
\begin{equation}
R^a_b=d\omega^a_b +\omega^a_c\wedge \omega^c_b
\end{equation}
where $T^a=\frac{1}{2}T^a_{bc} e^b\wedge e^c$ is the torsion 2-form 
and $R^a_b=\frac{1}{2}R^a_{bcd} e^c\wedge e^d$ is the curvature 2-form.
The Cartan identities read 
\begin{equation}
dT^a +\omega^a_b\wedge T^b =R^a_b\wedge e^b
\end{equation}
and
\begin{equation}
dR^a_b +\omega^a_c \wedge R^c_b -R^a_c\wedge \omega^c_b =0
\end{equation}
which is nothing but the Bianchi identity. 

In tensor components, 
\begin{equation}
T^\mu _{\alpha\beta}=\frac{1}{2} (\Gamma^\mu_{\alpha\beta}-\Gamma^\mu
_{\beta\alpha} ) 
\end{equation}
where 
\begin{equation}
\Gamma^\mu _{\alpha \beta}=\frac{1}{2}g^{\mu\nu} ( g_{\nu \beta , \alpha}
+g_{\nu \alpha , \beta} -g_{\alpha \beta , \nu } )
\end{equation}
are the Christoffel symbols, and 
\begin{equation}
R^a_b=\frac{1}{2} R^a_{b\mu \nu } dx^\mu \wedge dx^\nu 
\end{equation}
where
\begin{equation}
R^\alpha_{\beta \mu \nu } =\partial_\mu  \Gamma^\alpha_{\nu \beta } 
-\partial_\nu \Gamma^\alpha_{\mu \beta }
+\Gamma^\alpha_{\mu \gamma}\Gamma^\gamma_{\nu\beta}
-\Gamma^\alpha_{\nu \gamma}\Gamma^\gamma_{\mu\beta}
\end{equation}
is the Riemann tensor. As a simple example, consider the 2-sphere
\begin{equation}
ds^2=r_o^2(d\theta^2+\sin^2\theta d\phi^2) =(e^1)^2+(e^2)^2
\end{equation}
where $r_o$ is the radius of the sphere, and $e^1=r_od\theta$ and 
$e^2=r_osin\theta d\phi$. The structure equations reduce to 
\begin{equation}
de^1=-\omega^1_2\wedge e^2=0 \ \ \  \ and \ \ \ \ 
de^2=-\omega^2_1 \wedge e^1 =r_o cos d\theta \wedge d\phi 
\end{equation}
where $\omega^1_2 =-cos \theta d\phi$ is the connection 1-form. 
Note that $T^a=0$ here.
The curvature 2-form becomes 
\begin{equation}
R^1_2=R^1_{212}e^1\wedge e^2
\end{equation}
Note also that 
\begin{equation}
R^1_2 =d\omega^1_2 =\frac{1}{r^2} e^1\wedge e^2.
\end{equation}
The volume  form on $S^2$ is $\Omega =r_o^2\sin \theta d\theta \wedge d\phi$ with 
$\int \Omega =4\pi r_o^2$. 

Metrics can also be defined on the group manifolds like those of $SU(N)$.
A metric on $G$ is defined by the inner product 
\begin{equation}
(g'(0),h'(0))=-Tr (g^{-1}_o g'(0) g^{-1}_o h'(0))
\end{equation}
where $g(t)$ and $h(t)$ are two curves in $G$ which have an intersection 
at $t=0$:
\begin{equation}
g(0)=h(0)=g_o
\end{equation}
This metric is positive definite and multiplication on right or left 
correspond to isometries of the metric.  

Betti numbers ( $b_m$ ) 
can also be ascribed to the Riemannian manifolds. Compact orientable 
manifolds obey the \emph{Poincare' duality} which states that $H^p$ is 
dual to $H^{n-p}$. This implies $b_p=b_{n-p}$ which for a four-dimensional
manifold becomes $b_o=b_4$   and $b_1=b_3$. Note that $b_0=b_4$ counts 
the number of disjoint pieces  of ${\mathcal M}$, and 
$b_1=b_3$ vanishes if the manifold is simply connected. Recall that 
$\chi =b_o-b_1+b_2-b_3+b_4$ is the Euler characteristic of the manifold.
The following is an important theorem which relates the 
local curvature and the global characteristics of 
hypersurfaces in $R^n$:\\
 \emph{ Gauss-Bonnet Theorem:} For any compact, even-dimensional 
hypersurface $S$ in $R^{n+1}$, 
\begin{equation}
 \int_S \kappa =\frac{1}{2} \gamma_n \chi (S)
 \end{equation}
 where $\kappa$ is the curvature of $S$ ( see below ), $\chi (S)$ is the Euler 
 characteristic of $S$ and $\gamma_n$ is the volume of the unit 
$n-$sphere given by 
\begin{equation}
V_n=\frac{ 2(\pi )^{ \frac{n+1}{2} } }{ \Gamma (\frac{n+1}{2}) }
\end{equation}
where $\Gamma$ is the gamma function ( $\Gamma (1/2)=\sqrt{\pi}$, 
$\Gamma(1)=1$, and $\Gamma (x+1)=x\Gamma (x)$ ).

In a curved manifold, the Hodge * operation involves the metric determinant
$g$ and $\epsilon_{\mu\nu ...}=g\epsilon^{\mu \nu ...}$:
\begin{equation}
\ ^*(dx^{\mu_1} \wedge ... \wedge dx^{\mu_p} ) =
\frac{\sqrt{|g|}}{(n-p)!} \epsilon^{\mu_1...\mu_p}_{\mu_{p+1}
...\mu_n} dx^{\mu_{p+1}}\wedge ... \wedge dx^{\mu_n}
\end{equation}

For a scalar field $\phi$ on an $n-$dimensional Riemannian manifold, 
$\ ^*\phi=\sqrt{|g|}\phi dx^1\wedge...\wedge dx^n$. We also have
$\epsilon =\ ^*1$, $\ ^{**}T=(-1)^{k(n-k)}T$ for Euclidean signature 
and $\ ^{**}T=-(-1)^{k(n-k)} T$ for Minkowskian signature.

Torus $T^2=S^1\times S^1$ is a compact 2-dimensional manifold which can 
be covered by the coordinates $0\leq ( \theta_1 , \theta_2 )\leq 2\pi$.
A torus can not be covered with coordinates which are
globally smooth. For example, the $\theta_1$ and $\theta_2$ coordinates 
are discontinuous at the identified circles $\theta_i=0$ and 
$\theta_i =2\pi$. The 1-forms $d\theta_1$ and $d\theta_2$ are 
therefore not exact.  These two 1-forms provide a basis for the first 
cohomology group of $T^2$, with dim$ H^1 ( T^2) =2$. Also $H^2 (T^2 )$ 
has the basis $d\theta_1 \wedge d\theta_2$ and $dim H^2 ( T^2 ) =1$. We 
therefore have 
\begin{equation}
\sum^2_{p=0} (-1)^p dim H^p(T^2) = +1-2+1=0
\end{equation}
which is equal to the Euler characteristic of the torus.

Consider an oriented $n-$dimensional hyper-surface $X$ in $R^{n+1}$. 
 The outward pointing normal $\hat{n} (x)$ to 
 this hyper-surface maps $X$ into an $n$-sphere $S^n$ 
 and is called the \emph{ Gauss map}:
 \begin{equation}
 g:X\rightarrow S^n
 \end{equation}
 The Jacobian of this map $J_g (x)$ is called the curvature of $X$ at 
 $x$ and is denoted by $\kappa (x)$. For an $n$-sphere, 
 $\kappa =\frac{1}{r^n}$ independent of the point $x$ on $S^n$.\
\\ \ \\
{\bf \Large 12. Two dimensional ferromagnet} \\

Consider a three-component scalar field $\phi_a$ ( a=1,2,3 ) with the following
$O(3)$ Lagrangian density ( Belavin and Polyakov 1975, Rajaraman 1988 )
\begin{equation}
{\mathcal L} =\frac{1}{2}\partial^\mu \phi_a \partial_\mu \phi_a.
\end{equation}
The three components of the scalar field are constrained to the surface of 
a sphere :
\begin{equation}
\phi_a\phi_a =\phi_1^2+\phi_2^2+\phi_3^2 =1.
\end{equation}
Using the method of Lagrangian multipliers, the corresponding field equation
is found to be
\begin{equation}
\Box \phi_a -(\phi_b \Box \phi_b )\phi_a =0.
\end{equation}
Static configurations on the $xy$-plane are described by the following equation
\begin{equation}
\nabla^2 \phi_a -(\phi_b \nabla^2 \phi_b ) \phi_a =0
\end{equation}
Note that 
\[
\phi_a\phi_a =1 \ \ \ \rightarrow\ \ \  \phi_a\vec{\nabla} \phi_a =0 
\]
\[
\rightarrow \ \ \ \nabla .(\phi_a\vec{\nabla} \phi_a )=\vec{\nabla} \phi_a.
\vec{\nabla} \phi_a +\phi_a \nabla^2 \phi_a =0 
\]
\begin{equation} 
\rightarrow
\ \ \ \phi_a \nabla^2 \phi_a =-\vec{\nabla} \phi_a .\vec{\nabla} \phi_a
\end{equation}
The total energy of the system is given by
\begin{equation}
E=\int \frac{1}{2} \vec{\nabla} \phi_a.\vec{\nabla} \phi_a d^2x.
\end{equation}
We therefore have a $S^2$ fiber sitting on every point  of the 
$xy$-plane, forming a  bundle space which is locally $S^2\times R^2$. 
The classical vacuum of the system is at $\phi_a =constant$. $E$ vanishes 
for the vacuum. The $O(3)$ symmetry is spontaneously broken by the vacuum, 
which can be arbitrarily chosen to be at $(0,0,1)$. A finite-energy, 
localized solution of (210) is described by two functions $\phi_1 (x,y)$ 
and $\phi_2 (x,y)$ with $\phi_{1,2}\rightarrow 0$ as $r\rightarrow \infty$. 
The points on the $xy$-plane can therefore be transformed to 
a sphere using a stereographich map, in which all points on a circle with 
$r\rightarrow \infty$ are mapped to the  "north pole". This identification
is allowed so long as 
the single-valuedness of the $\phi_a$-field is 
concerned, since all points on this large circle reside at the same 
vacuum point $(0,0,1)$. In solid state theory, the $\phi_a$-field
may describe
the order parameter of a 2D ferromagnet.
The $S^2\rightarrow S^2$ mapping from the configuration
$S^2$ to the field $S^2$, 
therefore allows the following geometrical description of the 
system: 
\begin{itemize}
\item{Spin waves are sections of the bundle space which is now 
compactified to $S^2\times S^2$.}  
\item{Spin waves belong to 
distinct homotopy classes. The appropriate homotopy group 
is $\pi_2 (S^2)=Z$.}  
\item{Each mapping can be characterized by a 
winding number $n$ which counts the Brouwer degree of the map.  }
\end{itemize}

This 
winding number is obtained as 
\[
Q=\frac{1}{8\pi } \int \epsilon_{ij} \epsilon_{abc} 
\phi_a \partial_i \phi_b \partial_j \phi_c d^2x
\] 
\begin{equation}
= \frac{1}{8\pi } \int \epsilon_{ij} \epsilon_{abc}\phi_a
\frac{\partial \phi_b}{\partial \theta_l}\frac{\partial \theta_l}{\partial
x^i}\frac{\partial \phi_c}{\partial \theta_m}\frac{\partial \theta_m}{\partial x^j}
d^2 x \\ = \frac{1}{8\pi } \int \epsilon_{lm} \epsilon_{abc}
\phi_a \frac{\partial \phi_b}{\partial \theta_l}\frac{\partial \phi_c}{\partial
\theta_m } d^2 \theta 
\end{equation}
But $\frac{1}{2} \epsilon_{lm}\epsilon_{abc} \frac{\partial \phi_b}{\partial
\theta_l}\frac{\partial \phi_c}{\partial\theta_m} d^2\theta$ is 
the surface element of the internal $S^2$.  Therefore 
\begin{equation}
Q=\frac{1}{4\pi} 
\int d\vec{S}_{int}.\vec{\phi} =n
\end{equation}
Note that $d\vec{S}_{int}$ is parallel to the radius vector $\vec{\phi}$ of 
the unit sphere in the $(\phi_1, \phi_2, \phi_3)$ space. 

This interesting system can also be formulated on the complex plane, using
the stereographic projection
\begin{equation}
\omega =\frac{2\phi_1}{1-\phi_3}+i\frac{2\phi_2}{1-\phi_3}
\equiv \omega_1+i\omega_2
\end{equation}
The north pole of the $\phi$-sphere is projected to $|\phi |\rightarrow 
\infty$. Note that this projection is conformal (i.e. preserves 
angles). 

Consider the self-duality relation
\begin{equation}
\ ^*d\omega =-id\omega
\end{equation}
which leads to the equations
\begin{equation}
\frac{\partial \omega_1}{\partial x^1}=\pm \frac{\partial \omega_2}{\partial
x^2},
\end{equation}
 and 
 \begin{equation}
\frac{\partial \omega_1}{\partial x^2}=\mp \frac{\partial \omega_2}{
\partial x^1}.
\end{equation}
Since these are the familiar analyticity conditions, 
any analytic function $\omega (z)$ or $\omega (z^* )$, is  a 
solution of (216). It can be shown that such solutions in fact minimize 
the energy functional 
\begin{equation}
E=\int \frac{1}{2} \partial_i \phi_a \partial_i \phi_a d^2x
=\int \frac{|d\omega / dz|}{(1+\frac{1}{4}|\omega |^2 )^2}
\end{equation}
The topological charge  for these solutions is 
\begin{equation}
Q=\pm \frac{1}{4\pi}E.
\end{equation}
A simple solution with 
\begin{equation}
Q=\frac{1}{4\pi } E =n
\end{equation}
is given by 
\begin{equation}
\omega (z) =\frac{(z-z_o)^n}{\lambda^n}
\end{equation}
where $z_o$ and $\lambda$ are constants. Note that $Q$ and $E$ do not 
depend on $\lambda$ and $z_o$ which shows that the solutions can be 
scaled up or down or displaced on the $z$-plane.

It is seen that the nonlinear $O(3)$ model is quite simple, yet 
rich in structure. This simple model provides a good insight into 
the more complicated systems like the Yang-Mills instantons. 
\\ \ \\
{\bf \Large 13. Instantons in the $CP_N$ model}\\

Real projective spaces $P_N (R)$ are lines in $R^{N+1}$ which pass 
through the origin. For example, $P_3 (R) =SO(3)$.  $CP_N$ is the 
complex version of $P_N$.

Consider $N+1$ complex scalar fields $\phi_a$ ( $a=1,2,..., N+1$ ), 
with the following Lagrangian density (for more details, see Eichenherr 1978)
\begin{equation}
{\mathcal L} =\partial^\mu \phi^*_a \partial_\mu \phi_a 
+\phi^*_a ( \partial^\mu \phi_a ) \phi^*_b \partial_\mu \phi_b
\end{equation}
in which summation over $\mu$ and $a$ indices is implied. 
$x^\mu$ ( $\mu=1,2$ ) is considered to be the ( $xy$ ) Euclidean plane. 
The complex fields are subject to the constraint
\begin{equation}
|\phi_1|^2 +... +|\phi_{N+1}|^2 =1
\end{equation}
or $\phi^*_a\phi_a =1$. These complex fields form an $N$-dimensional 
complex projective ( $CP_N$ ) space. 
It is interesting to note that 
the Lagrangian density (223) is invariant under the local gauge 
transformation
\begin{equation}
\phi_a' (x)=\phi_a (x) e^{i\Lambda (x)}
\end{equation}
without a need to introduce any gauge potentials. A vector field 
$A^\mu (x)$, however, can be defined according to 
\begin{equation}
A^\mu =i\phi_a^* \partial^\mu \phi_a
\end{equation}
which is real since $\phi^*_a \partial^\mu \phi_a$ is pure imaginary. 
Using this definition, the Lagrangian density (223) can be written 
as 
\begin{equation}
{\mathcal L} =(D^\mu\phi_a )^* D_\mu \phi_a
\end{equation}
where
\begin{equation}
D_\mu \phi_a =(\partial_\mu +iA_\mu ) \phi_a 
\end{equation}
in resemblance to the covariant differentiation in electromagnetism.
Note, however, that $A^\mu$ is not a new degree of 
freedom, and it is determined solely in terms of the $\phi_a$ 
field. Localized solutions with finite energy should satisfy 
the following asymptotic behavior
\begin{equation}
\phi_a \rightarrow C_a e^{i\psi (\theta )}
\end{equation}
where $C_a$ are constants satisfying $C_a^*C_a =1$, and 
$\psi (\theta )$ is the common phase of the fields ( $\theta $ is the 
polar angle in the $xy$-plane ). This phase implies the winding 
number 
\begin{equation}
Q=\frac{1}{2\pi} \int d\psi =\frac{1}{2\pi } 
\oint \frac{d\psi }{d\theta } d\theta=n
\end{equation}
where $n$ is an integer. Similar to what we had in the case of cosmic 
strings, a $\pi_1 (S^1 )$ homotopy group is involved and solutions 
having different winding numbers belong to distinct homotopy classes. 
\\ \ \\
{\bf \Large 14. Skyrme model}\\

T.H.R. Skyrme  introduced a nonlinear model in 1960s, with soliton 
solutions approximately describing  baryons ( Skyrme, 1961 ).  
This model is 
based on the nonlinear sigma model, augmented with a nonlinear 
term in the Lagrangian which stabilizes the solitons. The Skyrme 
Lagrangian is therefore given by 
\begin{equation}
{\mathcal L}_{Sk}=-\frac{f_\pi^2}{4}Tr (U^\dagger \partial_\mu U U^\dagger
\partial^\mu U ) +\frac{1}{32\alpha^2 } Tr ([ U^\dagger \partial_\mu U
, U^\dagger \partial_\nu U ]^2 )
\end{equation}
where $U$ is the field described by a unitary matrix,  $\alpha$ is a
dimensionless coupling constant ( $\simeq 5$ ), and $f_\pi^2$ is the pion decay 
constant ( 130-190 MeV ). The pion mass term can also be included in (231) by adding 
the term $-\frac{1}{2}m_\pi Tr(U+U^\dagger )$.  
Soliton solutions of the Skyrme model are called \emph{ Skyrmions}. They 
have interesting topological properties, in relevance to 
the low energy properties of baryons. Witten (1983) showed that the 
Skyrme model is the high-$N_c$ limit of QCD, where $N_c$ is the number 
of colors. 
For a review of the mathematical 
developments in the Skyrme model, the reader is referred to Gisiger and 
Paranjape ( 1998 ). The geometrical aspects of the Skyrme model were 
first discussed by Manton and Ruback ( 1986 ). 

Since $U$ becomes constant at spatial infinity, the points in $R^3$ can
be mapped onto a $S^3$. Soliton solutions, therefore, define a mapping 
from  this $S^3$ to the group manifold of $SU(2)$ which is also an $S^3$.
These solutions can therefore be classified according to the third 
homotopy group $\pi_3$:
\begin{equation}
\pi_3 : S^3\longrightarrow S^3
\end{equation}
This homotopy group is isomorphic to the group of integers under addition,
and the topological charges label the distinct sectors.  The topological
current of the Skyrme model is given by
\begin{equation}
B^\mu =\frac{1}{24\pi^2}  \epsilon^{\mu \nu \alpha \beta} Tr (U^\dagger
\partial_\nu U U^\dagger \partial_\alpha U U^\dagger \partial_\beta U )
\end{equation}
with the corresponding charge identified with the baryon number:
\begin{equation}
B=\frac{1}{24\pi^2} \int d^3x \epsilon^{ijk} Tr (U^\dagger
\partial_i U U^\dagger \partial_j U U^\dagger \partial_k U ).
\end{equation}
 The simplest Skyrmion is obtained using the ansatz
 \begin{equation}
U=e^{i\hat{r}.\vec{\tau} f(r)}
\end{equation}
where $\vec{\tau}=(\tau_1,\tau_2,\tau_3)$ are the Pauli matrices, 
$\hat{r}$ is the unit vector in the radial direction, 
and $f(r)$ is a function to be determined by the field equation.  The boundary
conditions which are required to have  a well-defined solution are
\begin{equation}
f(0)=\pi \ \ \ \ and \ \ \ \ f(\infty )=0.
\end{equation}

The simplest description of the Skyrme model is in terms of unitary
$2\times 2$ matrices $U$ which belong to $SU(2)$.

Houghton et al. ( 1998 ) introduced a new ansatz for the Skyrme model
\begin{equation}
U(r,z)=exp(if(r)\hat{n}_R.\vec{\sigma} )
\end{equation}
where  
\begin{equation}
\hat{n}_R =\frac{1}{1+|R|^2} ( 2Re(R), 2Im(R), 1-|R|^2 )
\end{equation}
Note that the complex coordinate $z$ is related to the polar coordinates 
$\theta$ and $\phi$ via $z=tan(\theta /2) exp(i\phi )$ and $R(z)$ is 
a rational map $R(z) =p(z)/q(z)$, where $p$ and $q$ are polynomials of 
maximum degree $N$. The boundary conditions $f(0)=k\pi$ ( $k\in Z$ ) 
and $f(\infty )=0$ are implied. The degree of the rational map 
$R(z)$ determines the baryon number ( $B=Nk$ ). Therefore, all baryon 
numbers can be obtained with $k=1$, using the ansatz (237). The $N=1$ 
case is the Skyrme's original hedgehog ansatz. The ansatz (237) leads to 
the following expression for the total energy ( Houghton et al. 1998 )
\begin{equation}
E=\int \left[ \lambda_1^2 +\lambda_2^2+\lambda_3^2 +
(\lambda_1\lambda_2)^2 +(\lambda_2\lambda_3)^2 +(\lambda_1\lambda_3)^2 \right] d^3x
\end{equation}
where $\lambda_1^2$, $\lambda^2_2$, and $\lambda_3^2$ are the eigenvalues 
of the symmetric strain tensor 
\begin{equation}
D_{ij} =-\frac{1}{2}Tr ((\partial_i U U^{-1})(\partial_j U U^{-1})).
\end{equation}
The baryon number density is given by 
\begin{equation}
b=\frac{1}{2\pi^2} \lambda_1\lambda_2\lambda_3.
\end{equation}
It can be shown that 
\begin{equation}
\lambda_1=-f'(r), \ \ \ and  \ \ \ \lambda_2=\lambda_3=
\frac{\sin f}{r} \frac{1+|z|^2}{1+|R|^2} |\frac{dR}{dz}|
\end{equation}
A Bogomol'nye-type lower limit to the energy functional exists:
\begin{equation}
E\geq 4\pi^2 (2N+\sqrt{I})
\end{equation}
where 
\begin{equation}
I=\frac{1}{4\pi} \int \left( \frac{1+|z|^2}{1+|R|^2} 
|\frac{dR}{dz}| \right)^4 \frac{2idzd\bar{z}}{(1+|z|^2)^2}
\end{equation}
The energy limit can be written as $E\geq 12\pi^2 N$, 
since $I$ itself satisfies the inequality $I\geq N^2$.
\\ \ \\
{\bf \Large 15. Solitons and noncommutative geometry}\\

Noncommutative geometry provides a nice tentative framework  for the
unified description
of gauge fields and nonlinear scalar fields responsible for the 
spontaneous breakdown of gauge symmetries
(Connes 1985, Coquereaux et al. 1991, 
Chamseddine et al., 1993a, Madore 
1995, and Okummura et al. 1995 ). 

It has been shown that noncommutative geometry can also deal with 
gravity and leads to generalized theories of gravity like scalar-tensor
theories ( Chamseddine et al. 1993b ). Here, we 
briefly describe the general mathematical structure of the $Z_2$-graded
noncommutative geometry and its relevance to the localized soliton-like
solutions, closely following Teo and Ting ( 1997 ). The reader is 
referred to this paper for further details.

The Yang-Mills-Higgs theory is formulated using differential forms
on ${\mathcal M}\times Z_2$, where ${\mathcal M}$ is an $n-$dimensional
Euclidean space and $Z_2$ is the cyclic group of order two
\begin{equation}
Z_2=\{ e,r|r^2=e \}
\end{equation}
An explicit matrix representation of this group is 
\begin{equation}
\pi (e) = \left( \begin{array}{ll} 1 & 0 \\ 0 & 1 \end{array} \right)
, \ \ \ \ and \ \ \ \ \pi (r) = \left( \begin{array}{ll} 1 & 0 \\ 0 & -1 \end{array} \right)
\end{equation}
The algebra of complex functions on $Z_2$ is denoted by ${\mathcal M}_2^+$
and is a subalgebra of ${\mathcal M}_2$ which is the algebra generated 
by the Pauli matrices. An element of ${\mathcal M}^+_2 \otimes C$ is thus 
in the following form
\begin{equation}
\left( \begin{array}{ll} \mbox{$f_1$} & 0 \\ 0 & \mbox{$f_2$} \end{array} \right)
\end{equation}
where $f_1$ and $f_2$ are complex functions.

A generalized $p-$form on ${\mathcal M}\times Z_2$ looks like
\begin{equation}
\eta = \left( \begin{array}{ll} A+B & C \\ C' & A'+B' \end{array} \right)
\end{equation}
where $A$ and $A'$ are $p-$forms on ${\mathcal M}$ ( the 
horizontal $p-$forms ), $B$ and $B'$ are $(p-2)-$forms on ${\mathcal M}$, 
and $C$ and $C'$ are $(p-1)-$forms on ${\mathcal M}$. Note that 
$\eta$ is a composite $p-$form on ${\mathcal M}\times Z_2$, since 
a generalized $p-$form can be written as $\eta =a\times A$ where
$a$ is a $q-$form on $Z_2$ ( vertical part ) and $A$ is a $(p-q)-$form
on ${\mathcal M}$ ( horizontal part ). Even-degree forms on $Z_2$ are
diagonal matrices while those of odd degree are off-diagonal. Operations
on differential forms like exterior differentiation and Hodge * operation
can be extended to the $Z_2$ forms. 

The generalized connection 1-form on ${\mathcal M}\times Z_2$ has the 
form 
\begin{equation}
\omega =\mathcal{A} +\theta +\phi
\end{equation}
where $\mathcal{A}$ is the Yang-Mills connection ( horizontal part  of 
$\omega$ ), and $\theta +\phi$ is the vertical part associated with the
internal $Z_2$. The corresponding internal symmetry consists of  two 
global $U(1)$'s for the two elements of $Z_2$, denoted by $U(Z_2)$. 
Under $g\in U(Z_2)$, $\theta$ is the  gauge invariant 
Maurer-Cartan 1-form.  The curvature 2-form corresponding to the connection
(249) is 
\begin{equation}
\Omega =d\omega =i\omega \wedge \omega =F +D_H \phi +m^2 -\phi^2
\end{equation}
where $F=d_H {\mathcal A} +i{\mathcal A}\wedge {\mathcal A}$ is the 
Yang-Mills curvature and 
\begin{equation}
D_H \phi =d_H \phi +i{\mathcal A}\wedge \phi.
\end{equation}
Writing $\mathcal{A}$ and $\phi$ explicitly in their matrix forms
\begin{equation}
\mathcal{A} =\left( \begin{array}{ll} A & 0 \\ 0 & B \end{array} \right)
, \ \ \ \ \phi=\left( \begin{array}{ll} 0 & \mbox{$\phi$} \\ \mbox{$\phi^\dagger$} & 0 \end{array} \right)
\end{equation}
where $A$ and $B$ are ordinary 1-forms and $\phi$ is a 
complex scalar field, the curvature 2-form becomes
\begin{equation}
\Omega =\left( \begin{array}{ll} \mbox{$F+m^2-\phi \phi^\dagger $} & \mbox{$-D_H\phi$} 
\\ \mbox{$-D_H\phi^\dagger $} & \mbox{$G+m^2-\phi^\dagger \phi $} \end{array} \right)
\end{equation}
where $F=d_H A +iA\wedge A$, $G=d_H B + iB\wedge B$, and 
$D_H \phi = d_H \phi +i (A\phi -\phi B )$. The Euclidean action functional
then becomes
\[
S=\frac{1}{2}\int d^n x Tr (\Omega^{\dagger}_{ij} \Omega^{ij} ) 
\]
\begin{equation}
=\int d^n x \left\{ \frac{1}{2} F_{\alpha \beta} F^{\alpha \beta}  
+\frac{1}{2} G_{\alpha \beta } G^{\alpha \beta } 
+ 2D_\alpha \phi^\dagger D^\alpha \phi +2(m^2 -\phi^\dagger \phi )^2 \right\}
\end{equation}
Note that $D\Omega =0$ ( Bianchi identity ) and extremization of 
(254) leads to $D^*\Omega=0$ which can be expanded into the following equations
\[
D_\beta F_{\alpha \beta } =i(D_\alpha \phi^\dagger \phi -
\phi^\dagger D_\alpha \phi ) ,
\]
\[
D_\beta G_{\alpha \beta} =i(D_\alpha \phi^\dagger \phi -\phi^\dagger
D_\alpha \phi^\dagger ) ,
\]
and
\begin{equation}
D_\alpha D_\alpha \phi =-2 ( m^2 -\phi \phi^\dagger ) \phi .
\end{equation}
An element  of the composite symmetry group has the form $g=diag(g_1,g_2)$ where 
$g_1$ and $g_2$ belong to the Yang-Mills gauge group.  In the special 
case $g_1=g_2$ and $A=B$, the action functional reduces to 
\begin{equation}
S=\int d^n x \left\{ \frac{1}{4} F_{\alpha \beta} F^{\alpha \beta} 
+\frac{1}{2} D_\alpha \phi^\dagger D^\alpha \phi + \frac{\lambda}{2} 
( \phi^\dagger \phi )^2 -\mu^2 \phi^\dagger \phi \right\}
\end{equation}
where $\mu =em$, $\lambda =e^2$ and the rescaling $\omega \rightarrow 
e\omega$ has been implemented. If the gauge group responsible for the 
connection $A$ is $SU(2)$, this action will represent the ordinary 
Euclidean Yang-Mills theory coupled to a Higgs field with 
$F_{\alpha \beta} =\partial_\alpha A_\beta -\partial_\beta A_\alpha
+ ie[ A_\alpha , A_\beta ]$ and $D_\alpha \phi =\partial_\alpha \phi 
+ ie[A_\alpha , \phi ]$. The interesting aspect of the above formalism
is that the Higgs field $\phi$ responsible for the spontaneous symmetry
breaking of the Yang-Mills gauge field is now a gauge field associated 
with the internal $Z_2$ group. Moreover, the quadratic potential $V(\phi ,
\phi^\dagger )$ results naturally from the curvature of the connection
$\omega =\theta +\phi$ ( i.e. $\Omega =d\omega + i\omega \wedge \omega =
m^2 -\phi^2$, where only the vertical part has now been considered ). 

If the manifold ${\mathcal M}$ is one dimensional, the horizontal guage 
field $A$ dissappears since $A_\mu$ has only one component $\mu =1$ and 
the corresponding curvature vanishes. The Lagrangian density then reduces to 
\begin{equation}
{\mathcal L} =\frac{1}{2} |\frac{d\phi }{dx} |^2 +\frac{e^2}{2}
( |\phi|^2-m^2 )^2,
\end{equation}
which leads to the field equation 
\begin{equation}
\frac{d^2\phi}{dx^2} =2e^2\phi (|\phi|^2 -m^2).
\end{equation}
The ansatz $\phi (x) =\chi(x) exp(ikx)$ with real $\chi (x)$, leads to $k=0$ and 
\begin{equation}
\phi(x) =\chi(x) =\pm m\tanh (emx).
\end{equation}
This is the familiar kink ( anti-kink ) solution of the nonlinear 
Klein-Gordon equation in the static case.

The topological current $J=\frac{1}{2m}\frac{\partial \phi}{\partial x}$ 
leads to the charge $Q=\int Jdx=\pm 1$ for the kink ( anti-kink ). 
The kink ( anti-kink ) solution satisfies the 
generalized ( anti- ) self-duality relations 
\begin{equation}
\ ^* \Omega \wedge \theta =\pm im\Omega
\end{equation}
This self-duality condition also provides a lower limit for the 
soliton energy. This lower limit ( $E=\frac{4}{3}em^3$ ) is 
realized for the kink ( anti-kink ) solution. 

In a 2-dimensional Euclidean space with $U(1)$ as the 
underlying gauge symmetry (256) reduces to the static abelian Higgs 
model ( see equation 163 ). The string solution  
of Nielsen and Olesen fulfills the ( anti- ) 
self-duality relation
\begin{equation}
\ ^*\Omega =\pm \Omega
\end{equation}
similar to that of the non-abelian magnetic monopole. Note that 
here, $\Omega_{23} =-\frac{1}{m} D_2 ( Re \phi + iIm \phi \tau_3 )$, 
and $\Omega_{34}=\frac{e}{m^2} (m^2 -\phi^*\phi ) \tau_3$ 
where $B=F_{12}$ is the magnetic field, $\tau_3$ is the third Pauli 
matrix, and $D_\alpha$ is the $U(1)$ covariant derivative. 

In the $n=3$ case with $G=SU(2)$, the action  (256) can be 
identified with the static Yang-Mills-Higgs system which
possesses the monopole solutions of 't Hooft and Polyakov. The 
(anti) self-duality conditions (261) are now satisfied by the exact solutions
of Prasad and Sommerfield in the $\lambda \rightarrow 0$ limit.
\\ \ \\
{\bf \large Acknowledgements}\\

I would like to thank the Physics and Astronomy Department of the 
University of Victoria in B.C.
and in particular F. Cooperstock for hospitality during my sabbatical leave. 
Helpful suggestions from V. Karimipur and M. Hakim-Hashemi is gratefully
acknowledged. I also thank the  
Shiraz University Research Council for finantial support. \newpage
{\bf \Large References } \\ \ \\
Belavin, A.A., and Polyakov, A.M., {\it JETP Lett.}, {\bf 22}, 245, 1975.\\
Bogomol'nyi, E.B., {\it Sov. J. Nucl. Phys.}, {\bf 24}, 449, 1976.\\
Chamseddine, A.H., Felder, G., and Fr\"{o}hlich, J., {\it Nucl.
Phys.}, {\bf B395}, No. 3, 672, 1993a.\\
Chamseddine, A.H., Felder, G., and Fr\"{o}hlich, J., {\it Comm. Math. Phys.}, {\bf 155}, 205, 1993b.\\
Chern, S.S., {\it Complex Manifolds Without Potential Theory}, D. Van 
Nostrand Co., 1967.??\\
Coleman, S., {\it Classical Lumps and their Quantum Descendents}, Erice
Lecture (1975), in New Phenomena in Subnuclear Physics, Ed. A. Zichichi, 
Plenum, N.Y., 1977.\\
Connes, A., {\it Non-Commutative Differential Geometry},
{\it Publ. I.H.E.S.}, {\bf 62}, 41, 1985.\\
Coquereaux, R., Esposito-Farese, G., and Vaillant, G., {\it Nucl. Phys.},
{\bf B353}, 689, 1991.\\
Coquereaux, R., Esposito-Farese, G., and Scheck, F., {\it Int. J. Mod.
Phys.}, {\bf A7}, No. 26, 6555, 1992.\\
Darling, P.W.R., {\it Differential Forms and Connections}, Cambridge University
Press, USA, 1994.\\
Dijkgraaf, R., Witten, E., {\it Comm. Math. Phys.}, {\bf 129}, 393, 1990.\\
Drechsler, W., and Mayer, M.E., {\it Fiber Bundle Techniques in 
Gauge Theories}, Lecture notes in physics 67, Springer Verlag, Heidelberg, 1977.\\
Duan, Y., Xu, T., and Fu, L., {\it Prog. Theor. Phys.}, {\bf 101}, No. 2, 
467, 1999.\\
Eguchi, T., Gilkey, P., and Hanson, A.J., {\it Phys. Rep.}, {\bf 66}, 
No. 6, 213, 1980.\\
Eichenherr, H., {\it Nucl. Phys.}, {\bf B146}, 215, 1978.\\
Felsager, B., {\it Geometry, Particles, and Fields ( 2nd Ed. )}, Odense University 
Press., Gylling, 1983.\\
Gisiger, T., and Paranjape, M.B., {\it Phys. Rep. }, {\bf 306}, 109, 1998.\\
Goddard, P., and Olive, D., {\it Magnetic Monopoles in Gauge Field Theories},
{\it Reports in Progress in Physics}, {\bf 41}, 1357, 1978.\\
Guillemin, V., and Pollack, A., {\it Differential Topology}, Prentice-Hall
Inc., USA, 1974.\\
Henneaux, M., {\it Contemp. Math.}, {\bf 219}, ( AMS ), 1998.\\
Horva'thy, P.A., {\it Introduction to Monopoles}, Bibliopolis, Napoli, 1988.\\
Houghton, C.J., Manton, N.S., and Sutcliffe, P.M., {\it Nucl. Phys.}, 
{\bf B510}, No. 3, 507, 1998.\\
Henneaux, M., et al., {\it Comm. Math. Phys.}, {\bf 186}, 137, 1997.\\
Lamb, G.L., {\it Elements of Soliton Theory}, John Wiley and Sons,
New York, 1980.\\
Liu, F., and Mazenko, G.F., {\it Phys. Rev. Lett. }, {\bf 78}, 401, 1997.\\
Long, P.E., {\it An Introduction to General Topology}, 
Charles E. Merrill Pub. Co., U.S.A., 1971.\\
Madore, J., {\it An Introduction to Noncommutative Geometry and 
its Applications}, Cambridge University Press, Cambridge, England, 1995.\\
Manton, N.S., and Ruback, P.J., {\it Phys. Lett.}, {\bf B181}, 137, 1986.\\
Morris, M.S., and Thorne, K.S., {\it Am. J. Phys.}, {\bf 56}, 395, 1988.\\
Nambu, Y., {\it Lectures in Applied Math}, {\bf 21}, ( AMS ), 1985.\\
Nielsen, N.K., and Olesen, P., {\it Nucl. Phys. }, {\bf B61}, 45, 1973.\\
Nielsen, N.K., and Olesen, P., {\it Nucl. Phys. }, {\bf B291},No. 4, 829, 1987.\\
Okummura et al., {\it Prog. Theor. Phys. }, {\bf 94}, 1121, 1995.\\
Polyakov, A.M., {\it JETP Lett.}, {\bf 20}, 194, 1974.\\
Prasad, M.K., and Sommerfield, C.H., {\it Phys. Rev. Lett.}, {\bf 35}, 760, 1975.\\
Preskill, J., {\it Vortices and Monopoles}, Les Houches Lectures (1985), 
Caltech preprint  CALT-86-1287.\\
Rajaraman, R., {\it Solitons and Instantons}, Elsevier, The Netherlands, 
1987.\\
Riazi, N., and Gharaati, A.R., {\it Int. J. Theor. Phys.}, {\bf 37}, No. 3, 
1081, 1998.\\
Singer, I.M., and Thorpe, H.A., {\it Lecture Notes in Elementary Geometry 
and Topology}, Glenview, I11, Scott, Foresman, 1967.\\
Skyrme, T.H.R, {\it Proc. Roy. Soc. London}, {\bf 260}, 127, 1961.\\
Teo, E., and Ting, C., {\it Phys. Rev. }, {\bf D56}, No. 4, 2291, 1997.\\
T'Hooft, G., {\it Nucl. Phys.}, {\bf B79}, 276, 1974.\\
Vasheghani, A., and Riazi, N., {\it Int. J. Theor. Phys.}, {\bf 35},
587, 1996.\\
Yang, C.N., {\it Phys. Rev. Lett. }, {\bf 33}, No. 7, 445, 1974.\\                    
Yang, C.N., and Mills, R.L., {\it Phys. Rev. }, {\bf 96}, 191, 1954.\\
Witten, E., {\it Nucl. Phys.}, {\bf B223}, 433, 1983. \\
Wu, T.T., and Yang, C.N., {\it Phys. Rev.}, {\bf D12}, 3845, 1975.\\
\end{document}